\documentclass[reprint,amsmath,amssymb,aps,nofootinbib]{revtex4-2}
\usepackage{graphicx}
\usepackage{dcolumn}
\usepackage{bm}
\usepackage{physics}
\usepackage{hyperref}
\hypersetup{
     colorlinks=true,
     linkcolor=blue,
     filecolor=blue,
     citecolor = blue,      
     urlcolor=blue,
     }
\usepackage[super]{nth}
\usepackage{physics}

\usepackage[shortlabels]{enumitem}
\usepackage{bbm}
\allowdisplaybreaks

\newcommand{\hodge}{{\star}}
\newcommand{\extd}{\ensuremath{\text{d}}}

\newcommand{\bcovd}[2][]{\ensuremath{\bar{\nabla}_{#1}{#2}}}

\newcommand{\vbein}[3]{\ensuremath{e_{#2}^{({#1}){#3}}}}
\newcommand{\ivbein}[3]{\ensuremath{e_{\;\;\;\;\;{#3}}^{({#1}){#2}}}}
\newcommand{\gi}{\ensuremath{g^{(i)}_{\mu\nu}}}
\newcommand{\Wi}{\ensuremath{W^{(i)\mu}_{\;\;\;\;\;\;\nu}}}
\newcommand{\Gi}{\ensuremath{G^{(i)\mu}_{\;\;\;\;\;\;\nu}}}
\newcommand{\Tmunui}{\ensuremath{T^{(i)\mu}_{\;\;\;\;\;\;\nu}}}
\newcommand{\bvbein}[2]{\ensuremath{\bar{e}_{#1}^{\;\;{#2}}}}
\newcommand{\bivbein}[2]{\ensuremath{\bar{e}_{\;\;{#2}}^{{#1}}}}

\newcommand{\dx}{\ensuremath{\extd x}}

\newcommand{\Tcoeffs}{\ensuremath{T_{i_1\hdots i_D}}}

\begin{document}

\title{Black Holes in Multi-Metric Gravity}
\author{Kieran Wood}
\email{kieran.wood@nottingham.ac.uk}
\affiliation{School of Physics and Astronomy, University Of Nottingham, Nottingham NG7 2RD, UK}
\affiliation{Nottingham Centre Of Gravity, University Of Nottingham, Nottingham NG7 2RD, UK}

\author{Paul M. Saffin}
\email{paul.saffin@nottingham.ac.uk}
\affiliation{School of Physics and Astronomy, University Of Nottingham, Nottingham NG7 2RD, UK}
\affiliation{Nottingham Centre Of Gravity, University Of Nottingham, Nottingham NG7 2RD, UK}

\author{Anastasios Avgoustidis}
\email{anastasios.avgoustidis@nottingham.ac.uk}
\affiliation{School of Physics and Astronomy, University Of Nottingham, Nottingham NG7 2RD, UK}
\affiliation{Nottingham Centre Of Gravity, University Of Nottingham, Nottingham NG7 2RD, UK}

\begin{abstract}
    We construct a wide class of black hole solutions to the general theory of ghost free multi-metric gravity in arbitrary spacetime dimension, extending and generalising the known results in 4-dimensional dRGT massive gravity and bigravity. The solutions are split into three generic classes based on whether the metrics can be simultaneously diagonalised -- one of which does not exist in dRGT massive gravity nor bigravity, and is only possible when one has more than two interacting metric fields. We also linearise the general multi-metric theory to determine the dynamics of the massive spin-2 modes, including examples where this can be done analytically, and use the linear theory to discuss the stability of the 4-dimensional multi-Schwarzschild and multi-Kerr solutions. We explain how the instabilities that plague these solutions in dRGT massive gravity and bigravity carry across to the general multi-metric theory, touching upon ideas of dimensional deconstruction to make sense of the results.
\end{abstract}

\maketitle

\section{Introduction}\label{Sec:Intro}

Over the past decade or so, our understanding of the physics of interacting spin-2 fields has been revolutionised. It has long been known that general relativity (GR) is the unique local, two-derivative, nonlinear theory that describes a single, self-interacting, massless spin-2 field -- the graviton -- in four spacetime dimensions \cite{Gupta,Weinberg,Deser,Feynman_lecs,GR_from_QG}. For over a century, GR has remained our leading descriptor of the gravitational interaction, passing most observational tests to within a remarkable degree of precision. Despite its numerous successes, viable alternatives to GR have long been sought, as it is firstly a crucial task to develop theories of gravity for us to test GR against, and as there secondly remain multiple outstanding problems at the interface between gravity and particle physics (e.g. the nature of dark matter and dark energy, the cosmological constant problem, renormalisability etc.). One such alternative proposal, motivated originally by trying to explain the origin of the observed late-time accelerated expansion of the universe \cite{Riess,Perlmutter}, but later also by dark matter \cite{heavy_spin2_DM,bigravity_DM,DM_multigrav,Oscillating_DM} and the hierarchy problem \cite{ClockworkGrav,Deconstructing,ClockworkCosmo}, involves modifying gravity by introducing additional interacting spin-2 fields over and above the single massless graviton of GR. These theories then go by the helpful name of `multi-metric gravity', or simply, `multi-gravity'. There is, however, a no-go theorem, stating that theories that include multiple \emph{massless} interacting spin-2 fields are inconsistent \cite{no_interacting_massless_gravitons}; in these multi-metric theories, all but one of the spin-2 fields must be \emph{massive}. Therefore, any consistent multi-metric theory must be built off the back of a similarly consistent theory of massive gravity, which itself has a colourful history.

The story of massive gravity dates back to 1939, when Fierz and Pauli first wrote down the linearised theory describing a massive, self-interacting spin-2 field \cite{FierzPauli}. For a long while, it was the pervasive view that any nonlinear completion of the linear Fierz-Pauli theory would be pathological, owing to the emergence of the so-called Boulware-Deser (BD) ghost -- a problematic scalar mode equipped with wrong-sign kinetic term, signalling an instability of the vacuum -- as soon as nonlinear interactions were taken into account \cite{BD_Ghosts,BD_ghost_explicit}. However, the original BD analysis did not take into account all possible interaction terms, and a breakthrough finally came much later in 2010 when a satsifactory nonlinear theory of massive gravity was indeed constructed \cite{dRGT_1,dRGT_2,first_action_metric_form} and subsequently proved to be free of the BD ghost \cite{ghost_freedom_flat_ref,ghost_freedom_general_ref,ghost_freedom_stuckelberg,hamiltonian_analysis,Kluson,Kluson_note,Covariant_approach_no_ghosts}. The theory, built upon groundwork laid earlier in \cite{EFTforMGs,Creminelli}, is now known as dRGT massive gravity, after its progenitors: de Rham, Gabadadze and Tolley (there were important contributions also by Hassan and Rosen \cite{first_action_metric_form,ghost_freedom_flat_ref,ghost_freedom_general_ref,ghost_freedom_stuckelberg}). In four spacetime dimensions, it describes the five degrees of freedom of a single propagating massive graviton via a framework involving interactions between the physical spacetime metric and an auxiliary reference metric, which one inserts by hand (typically taken to be Minkowski, though this need not be the case). By providing a kinetic term for the reference metric, thereby promoting it to a second dynamical field, one obtains the theory of bigravity \cite{HR1}, which, due to the special structure of the dRGT interactions, is also ghost free \cite{HR2}. The generalisation to multiple interacting metric fields followed soon after in \cite{interacting_spin2}, although the general multi-metric theory is only devoid of the BD ghost up to certain conditions, upon which we shall elaborate in section \ref{Sec:Review}. For further details regarding the development and phenomenology of these theories, we refer the reader to the excellent and comprehensive reviews \cite{dR_review,Hinterbichler_review} on massive gravity, as well as \cite{bigravity_review} on bigravity.

Armed with a consistent ghost free framework for multi-metric gravity, the natural next step is to begin to search for physical solutions that might describe our world. Much work has focussed on such solutions in the realm of cosmology, with a lot of initial excitement surrounding the potential of the theories to address the dark energy problem (see any of the reviews cited above and references therein for details). However, work has also steadily been ongoing to construct black hole solutions in these theories and understand their properties; such an endeavour is of course crucial to test the theory, since these objects do exist in our universe and provide a natural arena to look for deviations from GR \cite{GR_tests_astro}.

In the multi-metric theory, the simplest background solutions are those in which each metric is proportional, via a constant conformal factor, to some common GR background \cite{consistent_spin2,prop_bg_multigrav}. Clearly, these proportional solutions then contain multi-metric analogues of all of the known GR black hole solutions. In 4-dimensional dRGT massive gravity and bigravity, however, additional non-proportional solutions are known to exist in which each metric individually is patterned as a GR solution, but the two metrics are not simultaneously diagonal \cite{KoyamaNiz,Exact_schdS,dR_BHs,horizon_struct_bigravity,spherical_sym_sols,charged_BHs_bigravity,rotating_BHs_bigravity,Rotating_AdS_bigravity}. Furthermore, owing to the absence of the GR no-hair theorems \cite{no_hair_1,no_hair_2,no_hair_3} in massive gravity, these theories contain solutions that are known only numerically and are completely foreign to GR, in which the black holes are endowed with a cloud of massive graviton hair \cite{Hairy_BH_AdS,Hairy_BHs_flat}. 

The linear stability of a number of these dRGT and bigravity black holes has also been studied, with some intriguing results. In \cite{GL_instability_bigravity,Kerr_instability_bigravity}, it was shown that the solution where both metrics are simultaneously proportional to the ($D=4$) Schwarzschild(-dS) metric suffers from a radial instability that is present for certain values of the graviton mass. The instability takes the same form as the Gregory-Laflamme instability afflicting 5-dimensional black strings \cite{GL_instability,Charged_GL_instability,AdS_GL_instability}; we shall see later on why this result should not be surprising. The proportional Kerr solution also suffers from this radial instability \cite{Kerr_instability_bigravity}, as well as a superradiant one that is again dependent on the size of the graviton mass \cite{Kerr_instability_bigravity,superradiance_ULS2,superradiance_massive_spin2,backreaction}. On the other hand, the non-proportional Schwarzschild solution appears to be linearly stable \cite{GL_unified,Stability_nonbidiag}, although whether or not it excites a (non-BD) ghost is unknown. Likewise, the final state of any of the aforementioned instabilities of the proportional solutions remains to be determined.

The above results -- which are summarised nicely in the review \cite{BHs_review} -- as we stated, are known only for dRGT massive gravity and bigravity, which contain respectively one and two dynamical metric fields. It remains an open question as to how the relevant black hole physics carries over to the general multi-metric theory. In this work, we close this gap somewhat, by explicitly constructing a wide variety of black hole solutions to the general multi-metric theory (with both classes of allowed interaction structures -- see section \ref{Sec:Review}) in an arbitrary number of spacetime dimensions, generalising and extending the known dRGT/bigravity results in $D=4$. We also discuss the stability of the proportional solutions in the $D=4$ multi-metric theory, determining how the Schwarzschild and Kerr instabilities manifest when there are multiple massive spin-2 fields, rather than just one. The extensions to the multi-metric theory are natural, in the sense that many of the bigravity results carry over in the manner one might naively expect, although there are some subtleties regarding the non-proportional solutions that do arise in the multi-metric theory that are not present in dRGT/bigravity. Throughout, we construct a number of explicit example black hole spacetimes to illustrate these points, as well as tie in to ideas of dimensional deconstruction \cite{deconstructing_dims,Deconstructing} to make clear the link between the instabilities of multi-metric black holes and higher dimensional black strings. 

The structure of the paper, then, is as follows: in section \ref{Sec:Review} we review the general multi-metric theory in arbitrary spacetime dimension, and introduce its metric and vielbein formulations that will both aid calculations later; in section \ref{Sec:BG} we construct a wide array of arbitrary dimension background black hole solutions in each of the proportional, non-proportional and (new for multi-gravity) partially proportional branches; in section \ref{Sec:linear} we linearise the general multi-metric theory to express explicitly the dynamics of the spin-2 mass modes, providing examples where one may do this analytically; in section \ref{Sec:stability} we use this linear theory to extend the dRGT/bigravity results regarding the instability of the proportional Schwarzschild and Kerr solutions to the general multi-metric theory; finally we conclude in section \ref{Sec:conclusion}.

We work with natural units $c=\hbar=G=1$ throughout, and always use a mostly-plus metric signature.

\section{Review of multi-metric gravity}\label{Sec:Review}

The action for multi-metric gravity, living on some $D$-dimensional spacetime manifold $\mathcal{M}_D$, can be written conveniently in the vielbein formalism as a sum of $N$ Einstein-Hilbert kinetic terms \cite{EH_uniqueness_1,EH_uniqueness_2,EH_uniqueness_3} together with an interaction potential of degree $D$ coupling the various basis 1-forms (see e.g. \cite{interacting_spin2,ClockworkGrav,ClockworkCosmo,Cosmo_spin_2}):
\begin{align}\label{MultigravAction}
    S &= S_K + S_V + S_M
    \\
    S_K &= \sum_{i=0}^{N-1} \frac{M_i^{D-2}}{2} \int_{\mathcal{M}_D} R^{(i)}_{ab} \wedge \hodge^{(i)} e^{(i)ab}\label{MultigravKinetic}
    \\
    S_V &= -\sum_{i_1\hdots i_D=0}^{N-1} \int_{\mathcal{M}_D} \varepsilon_{a_1\hdots a_D} T_{i_1\hdots i_D} e^{(i_1)a_1}\wedge\hdots\wedge e^{(i_D)a_D}\label{MultigravPotential} \; .
\end{align}
The tetrad basis 1-forms are $e^{(i)a}=\vbein{i}{\mu}{a} \dx^\mu$, where the indices run from 0 to $D-1$, with the vielbeins defined through $\gi = \vbein{i}{\mu}{a} \vbein{i}{\nu}{b} \eta_{ab}$, and the shorthand $e^{(i)ab\hdots}$ in the kinetic term means $e^{(i)a}\wedge e^{(i)b}\wedge\hdots$. We say that the $(i)$ labels refer to a particular `site' within the interaction structure; indices are then raised/lowered site-wise: Latin indices with $\eta_{ab}$ and Greek indices with $\gi$, while we can swap between Latin and Greek indices using the vielbeins, via changes of basis. $R^{(i)}_{ab}$ is the curvature 2-form associated with the $i$-th (Levi-Civita) connection, with one index lowered by $\eta_{ab}$, so that the kinetic term is just $N$ copies of the usual Einstein-Hilbert action, written in the convenient language of differential forms. The $T_{i_1\hdots i_D}=T_{(i_1\hdots i_D)}$ are symmetric coefficients that characterise the interactions between the tetrads. Finally, $S_M$ is the action for the collective matter fields coupled to the theory.

We note that since we are working in an arbitrary number of dimensions, in principle the higher dimensional Lovelock invariants (the Euler-Poincaré forms, in differential form language) may also be included in the kinetic term $S_K$ \cite{Lovelock,Lovelock_bigrav}, although we restrict ourselves to only include the Einstein-Hilbert term for simplicity.

One can also express the theory described by Eq. \eqref{MultigravAction} in the more commonly used metric formalism \cite{first_action_metric_form,Multi_via_massive,metric_multigravity}, where the interaction term is instead written in terms of the characteristic building block matrices
\begin{equation}
    S_{i\rightarrow j} = \sqrt{g_{(i)}^{-1}g_{(j)}} \; ,
\end{equation}
with the matrix square root defined in the sense that $(S^2_{i\rightarrow j})^\mu_{\;\nu}=g^{(i)\mu\lambda}g^{(j)}_{\lambda\nu}$. Explicitly, the metric version of the potential term is:
\begin{equation}\label{metric potential}
    S_V = -\sum_{i,j}\int \dd^D x \sqrt{-\det g_{(i)}} \sum_{m=0}^{D} \beta_m^{(i,j)} e_m(S_{i\rightarrow j}) \; ,
\end{equation}
where the $\beta_m^{(i,j)}=\beta_m^{(j,i)}$ are arbitrary constants related to the $T_{i_1\hdots i_D}$ of the vielbein formalism in a manner we shall soon specify, and the $e_m(S)$ are elementary symmetric polynomials of the eigenvalues of $S$, given by:
\begin{align}
        e_0(\lambda_1,\lambda_2,...,\lambda_{D}) &= 1 \\
        e_1(\lambda_1,\lambda_2,...,\lambda_{D}) &= \sum_{1\leq i \leq D} \lambda_i \\
        e_2(\lambda_1,\lambda_2,...,\lambda_{D}) &= \sum_{1\leq i<j\leq D} \lambda_i \lambda_j \\
        \vdots\nonumber \\
        e_k(\lambda_1,\lambda_2,...,\lambda_{D}) &= \sum_{1\leq j_1<j_2<...<j_k\leq D} \lambda_{j_1}...\lambda_{j_k} \\
        \vdots\nonumber \\
        e_{D}(\lambda_1,\lambda_2,...,\lambda_{D}) &= \lambda_1 \lambda_2 ... \lambda_{D} \, .
\end{align}
They can also be constructed iteratively in terms of the trace of $S$, as:
\begin{equation}\label{sym pols}
    e_m(S)=-\frac{1}{m}\sum_{n=1}^m(-1)^n\Tr(S^n)e_{m-n}(S) \; .
\end{equation}
The structure of the building block matrices means that $S_{i\rightarrow j}=S_{j\rightarrow i}^{-1}$, so there is a sense in which the interactions are \emph{oriented} \cite{prop_bg_multigrav,SC_and_graph_structure}: we say that a term in the potential, Eq. \eqref{metric potential}, which contains $S_{i\rightarrow j}$ (\emph{not} $S_{j\rightarrow i}$) is positively oriented with respect to the $i$-th metric and negatively oriented with respect to the $j$-th metric. The orientation will affect the equations of motion for the $i$-th and $j$-th metrics differently, as we will soon see, and it is accounted for in the vielbein formalism within the structure of the $\Tcoeffs$.

These metric interactions, as they must, take precisely the special dRGT form that is required to remove the Boulware-Deser ghost \cite{dRGT_1,dRGT_2,first_action_metric_form}. It was argued in \cite{vielbein_to_rescue} that the vielbein theory described by the action \eqref{MultigravAction} is therefore ghost free only if it has an equivalent description in metric form, which happens whenever the so-called Deser-van Nieuwenhuizen symmetric vielbein condition,
\begin{equation}\label{SymmetricVierbeinCondition}
    e_{\;\;\;\;\;a}^{(i)\mu}e_\mu^{(j)b} = e^{(i)\mu b}e^{(j)}_{\mu a} \; ,
\end{equation}
is satisfied. The known vielbein models that satisfy this condition are those involving only pairwise interactions with no cycles (a cycle is e.g. $1\rightarrow2\rightarrow3\rightarrow1$), so such models are \emph{manifestly} ghost free \cite{ghost_freedom_multigravity,cycles,prop_bg_multigrav}, though more recently constructions that evade the arguments of \cite{vielbein_to_rescue} yet nevertheless remain ghost free were given in \cite{eff_matter_coupling_completion,beyond_pairwise_couplings}.

Both the metric and vielbein formalisms of multi-metric gravity prove useful calculational tools in different situations: indeed, we shall employ both at different points in this work, depending on which is more appropriate. When working with the vielbein formalism, we shall choose, however, to restrict to those ghost free vielbein theories that do permit a metric description, for simplicity. In particular, we shall take `chain' type interactions \cite{prop_bg_multigrav}, where the $i$-th metric interacts only with its nearest neighbours, and the interactions are always positively oriented from $i$ to $(i+1)$\footnote{In principle, `star' type interactions: where many metrics all couple to some common central metric but not to each other, are also allowed. The most general ghost free interaction one could choose in multi-gravity (at least, from those vielbein theories permitting a metric description) then consists of arbitrary combinations of `star' and `chain' type interactions (see \cite{prop_bg_multigrav} for more details). We will look at how our calculations in the later sections change for a star type interaction in appendix \ref{App:star}.}. Such a choice is natural if one thinks of the theory as arising from some sort of dimensional deconstruction \cite{deconstructing_dims,Deconstructing,ClockworkCosmo}. In terms of the $T_{i_1\hdots i_D}$, this means that one can only permit terms of the form $T_{iiii\hdots}$, $T_{i+1,iii\hdots}$, $T_{i-1,iii\hdots}$, $T_{i+1,i+1,ii\hdots}$ and so on.

Amongst the general class of theories described by Eq. \eqref{MultigravAction}, a particular model is specified entirely by a choice for both the number of metrics $N$ and the $T_{i_1\hdots i_D}$ coefficients. For chain type interactions, one can always neatly parametrise the $\Tcoeffs$ in terms of some new constants $\beta_m^{(i,i+1)}$, which characterise the interactions between the $i$-th and $(i+1)$-th metrics, as \cite{ClockworkCosmo}
\begin{align}
    D!\,T_{iiii\hdots i} &= \beta_0^{(i,i+1)} + \beta_D^{(i-1,i)}\label{betas1} \\
    D!\,T_{\{i+1\}^m\{i\}^{D-m}} &= \beta_m^{(i,i+1)}\label{betas2} \; ,
\end{align}
where $i=0,\hdots,N-1$ and $m=0,\hdots,D$, with $\beta_m^{(-1,0)}=\beta_m^{(N-1,N)}=0$ since on each of the two boundaries of the interaction chain the corresponding metrics have only one nearest neighbour. The factor of $D!$ is included so that these $\beta_m^{(i,i+1)}$ are then precisely the same $\beta_m^{(i,j)}$ as in the metric formalism Eq. \eqref{metric potential}. The sense of interaction orientation is encoded in $T_{iiii\hdots i}$, where positively oriented interactions contribute a $\beta_0$ term while negatively oriented interactions contribute a $\beta_D$ term. In practice, we will often choose to write $\beta_m^{(i,i+1)}=\alpha_i\beta_m$, where $\alpha_{-1}=\alpha_{N-1}=0$ but the rest of the $\alpha_i$ are equal, restricting to the case where the interactions between all of the metrics in the chain are characterised by the same set of parameters $\beta_m$. This is both to avoid having an abundance of free parameters in the theory and because, again, such a choice is natural from the deconstruction perspective. However, we shall keep the generic set of $\beta_m^{(i,i+1)}$ in for most of our expressions in order to be as general as possible.

The equations of motion for the general $D$-dimensional theory are:
\begin{equation}\label{Einstein eqs}
    M^{D-2}_{i} \Gi + \Wi = \Tmunui \; ,
\end{equation}
where $\Tmunui$ are the energy-momentum tensors for the various sites, and the new term $W$ characterises the effect of the interactions over and above the standard GR interactions. Explicitly, in vielbein form, $\Wi$ reads \cite{ClockworkCosmo} (see appendix \ref{App:eqs} for the derivation):
\begin{equation}\label{W tensor}
\begin{split}
    \Wi &= D!\vbein{i}{\nu}{a}\ivbein{i}{\mu}{[a}\ivbein{i}{\lambda_1}{b_1}\hdots\ivbein{i}{\lambda_{D-1}}{b_{D-1}]} \\&\times \sum_{j_1\hdots j_{D-1}} \mathcal{P}(i) T_{ij_1\hdots j_{D-1}} \vbein{j_1}{\lambda_1}{b_1}\hdots\vbein{j_{D-1}}{\lambda_{D-1}}{b_{D-1}} \; ,
\end{split}
\end{equation}
with $\mathcal{P}(i)$ counting the number of times the index $(i)$ appears in the interaction coefficients i.e. a term with $T_{ij_1\hdots j_{D-1}}$ has $\mathcal{P}(i)=1$, a term with $T_{iij_2\hdots j_{D-1}}$ has $\mathcal{P}(i)=2$, and so on. The equivalent metric formalism expression is:
\begin{equation}\label{metric eqs}
\begin{split}
    \Wi &= \sum_j \sum_{m=0}^D(-1)^m\beta_m^{(i,j)}Y_{(m)\nu}^\mu(S_{i\rightarrow j})
    \\
    &+\sum_k\sum_{m=0}^D (-1)^m\beta_{D-m}^{(k,i)}Y_{(m)\nu}^\mu (S_{k\rightarrow i}^{-1}) \; ,
\end{split}
\end{equation}
where (with respect to the $i$-th metric) $j$ denote positively oriented interactions, $k$ denote negatively oriented interactions, and we define
\begin{equation}\label{Y_def}
    Y_{(m)}(S) = \sum_{n=0}^m (-1)^n S^{m-n}e_n(S) \; .
\end{equation}

The $\Tmunui$ are not completely arbitrary: in order to remain ghost free, the forms of matter coupling that one can permit are severely restricted. In general, one must couple entirely \emph{separate} matter sectors to separate vielbeins, otherwise the BD ghost is resurrected \cite{ghost_doubly_coupled,on_matter_couplings,ghosts_matter_couplings_rev}. There is the notable exception, however, where a \emph{single} matter source can be coupled to multiple vielbeins in a ghost free manner through the special `effective' vielbein considered in \cite{on_matter_couplings,ghost_freedom_eff_metric,matter_coupling_multigravity,generalised_matter_couplings,eff_matter_coupling_completion}.

The non-interacting theory possesses $N$ copies of diffeomorphism invariance. Turning on the interactions, these diffeomorphisms are broken to a single surviving diagonal subgroup, so the theory propagates a single massless graviton (invariant under transformations of this subgroup) and $N-1$ massive gravitons, which are linear combinations of the original metric perturbations.

As a result of the Bianchi identities for each Einstein tensor, as well as the surviving diagonal diffeomorphism invariance, the $W$-tensor is subject to the \emph{Bianchi constraint} \cite{ClockworkCosmo}:
\begin{equation}\label{W constraint}
    \sum_{i=0}^{N-1} \abs{e^{(i)}}\nabla^{(i)\mu}W^{(i)}_{\mu\nu} = 0 \; .
\end{equation}
Whenever matter couples to one site only, or when there is no matter coupling at all, the divergences of each $W$-tensor (i.e. each term in the above sum) must instead vanish \emph{individually}, telling us that there can be no flow of energy-momentum across the chain of interactions.

\section{Background Solutions and Black Holes}\label{Sec:BG}

General solutions of the multi-metric theory can deviate markedly from solutions of GR, although it has long been known that the simplest multi-metric solutions are those where all metrics are proportional to some common GR background \cite{consistent_spin2,prop_bg_multigrav}. Such solutions are useful, because they tell us about those multi-metric solutions that are close to what we already know from GR. They also admit a sensible perturbative description that allows us to analyse the mass spectrum (indeed, we shall see this in section \ref{Sec:linear}). However, these simple solutions are also restrictive: viable cosmological solutions, for example, necessarily do not lie in this class \cite{viable_cosmologies,ClockworkCosmo} and, as we shall demonstrate shortly, there exist numerous black hole solutions where the metrics are not proportional (a fact that has been known in bigravity for some time \cite{charged_BHs_bigravity,rotating_BHs_bigravity,BHs_review}). Nevertheless, both types of solution are important, so we shall consider each in turn, utilising both metric and vielbein formalisms where appropriate to aid in their construction.

\subsection{Proportional Solutions}\label{Sec:prop sols}

We look for solutions to the multi-metric equations \eqref{Einstein eqs} where the metrics are conformally related to one another,
\begin{equation}\label{propBG}
    \gi = a_i^2 \bar{g}_{\mu\nu} \; ,
\end{equation}
where at this stage $\bar{g}_{\mu\nu}$ is some arbitrary fixed metric common to all sites. Since all of the metrics live on the same manifold $\mathcal{M}_D$, one only has the freedom to rescale the coordinates to fix \emph{one} of the $a_i$, so their values are physical, up to an overall normalisation.

With the ansatz \eqref{propBG}, the $i$-th vielbein and its inverse are given by:
\begin{align}
    \vbein{i}{\mu}{a} &= a_i\bar{e}_\mu^{\;\;a} \; ,
    \\
    \ivbein{i}{\mu}{a} &= \frac{1}{a_i}\bar{e}^\mu_{\;\;a} \; .
\end{align}
Therefore, one may express the $W$-tensor in terms of $\bar{e}_\mu^{\;a}$ only, and so remove the vielbeins from the sum over the $j$'s entirely:
\begin{equation}
\begin{split}
    \Wi &= D!\bvbein{\nu}{a}\bivbein{\mu}{[a}\bivbein{\lambda_1}{b_1}\hdots\bivbein{\lambda_{D-1}}{b_{D-1}]} \bvbein{\lambda_1}{b_1}\hdots\bvbein{\lambda_{D-1}}{b_{D-1}}\\&\times \sum_{j_1\hdots j_{D-1}} \mathcal{P}(i) T_{ij_1\hdots j_{D-1}} \frac{a_{j_1}\hdots a_{j_{D-1}}}{a_i^{D-1}} \; .
\end{split}
\end{equation}

Since all the vielbeins now belong to the same background geometry, they can be contracted. It is simple enough to show that the vielbeins contract to:
\begin{equation}
    D!\bvbein{\nu}{a}\bivbein{\mu}{[a}\bivbein{\lambda_1}{b_1}\hdots\bivbein{\lambda_{D-1}}{b_{D-1}]} \bvbein{\lambda_1}{b_1}\hdots\bvbein{\lambda_{D-1}}{b_{D-1}} = (D-1)!\delta^\mu_\nu \; ,
\end{equation}
and so the $W$-tensor takes the following simple, diagonal form, \emph{irrespective of} the precise identity of $\bar{g}_{\mu\nu}$:
\begin{equation}\label{W simple}
    \Wi = (D-1)!\delta^\mu_\nu \sum_{j_1\hdots j_{D-1}} \mathcal{P}(i) T_{ij_1\hdots j_{D-1}} \frac{a_{j_1}\hdots a_{j_{D-1}}}{a_i^{D-1}} \; .
\end{equation}

Taking chain type interactions, using the symmetry of the $T_{i_1\hdots i_D}$ coefficients, and parametrising them as in Eqs. \eqref{betas1} and \eqref{betas2}, one finds that the components of the $W$-tensor are explicitly given by:
\begin{equation}\label{W_comps_prop}
    \begin{split}
        \Wi = \delta^\mu_\nu\Bigg[& \sum_{m=0}^{D} \beta_m^{(i,i+1)} \binom{D-1}{m}a_{i+1}^m a_{i}^{-m} \\  &+\sum_{m=0}^{D} \beta_{D-m}^{(i-1,i)} \binom{D-1}{m} a_{i-1}^m a_{i}^{-m} \Bigg] \; .
    \end{split}
\end{equation}

The Bianchi constraint Eq. \eqref{W constraint} forces all the $a_i$ to be constant \cite{consistent_spin2}. Lowering an index in Eq. \eqref{W_comps_prop} tells us that $W^{(i)}_{\mu\nu}\propto \gi$; therefore, we should interpret our $W$-tensors as effective cosmological constants on each site $(i)$, which arise due to the interactions. The downstairs index Einstein tensor is scale invariant, that is, $G_{\mu\nu}(g)=G_{\mu\nu}(ag)$, so we have that $\Gi=\bar{G}^\mu_{\;\nu}/a_i^2$. We can absorb these factors of $a_i$ into the definitions of the effective cosmological constants and energy-momentum tensors so that the multi-metric equations reduce simply to $N$ copies of the standard Einstein equations for $\bar{g}_{\mu\nu}$:
\begin{equation}\label{vac_eqs}
    \bar{G}^\mu_{\;\nu} + \Lambda_i \delta^\mu_\nu = M_i^{-(D-2)}\bar{T}^{(i)\mu}_{\;\;\;\;\;\;\nu} \; ,
\end{equation}
where the effective cosmological constants $\Lambda_i$ are precisely the contributions from $\Wi$. Explicitly, the appropriate rescaled quantities are
\begin{align}
    \Lambda_i\delta^\mu_\nu &= \frac{a_i^2}{M_i^{D-2}}\Wi\label{CCbars} \; ,
    \\
    \bar{T}^{(i)\mu}_{\;\;\;\;\;\;\nu} &= a_i^2 \Tmunui\label{Tbars} \; ,
\end{align}
so that all terms in Eqs. \eqref{vac_eqs} now behave as if they lived on the \emph{common} background $\bar{g}_{\mu\nu}$ i.e. have their indices manipulated with $\bar{g}_{\mu\nu}$.

We can take differences of Eqs. \eqref{vac_eqs} to show that the effective cosmological constants must satisfy:
\begin{equation}
    \Lambda_i = \bar{\Lambda} \;\;\;  \forall \; i \; ,
\end{equation}
and similarly the energy-momentum tensors must all be proportional \cite{consistent_spin2},
\begin{equation}
    \bar{T}^{(i+1)\mu}_{\;\;\;\;\;\;\;\;\;\;\nu} = \left(\frac{M_{i+1}}{M_i}\right)^{D-2} \bar{T}^{(i)\mu}_{\;\;\;\;\;\;\nu} \; .
\end{equation}

This restriction on the matter sources may not be necessarily realistic in general, but it is certainly true in vacuum, where all the energy-momentum tensors vanish. Therefore, these proportional solutions describe perfectly acceptable vacua at the background level in multi-metric gravity. We shall construct some examples shortly, though before we get there, we note that the condition on the $\Lambda$'s is not as simple as it may first seem, since on the `boundary' sites of the interaction chain (i.e. $i=0$ and $i=N-1$) one set of $\beta$'s vanishes, so only one sum is present in the corresponding $W$-tensor. If we denote $\Lambda^{(+)}_i$ as $a_i^2/M_i^{D-2}$ times the sum involving $a_{i+1}$ and $\Lambda^{(-)}_i$ as $a_i^2/M_i^{D-2}$ times the sum involving $a_{i-1}$, then what we actually have is the following:
\begin{align}
    \Lambda^{(+)}_i+\Lambda^{(-)}_i &= \bar{\Lambda}  \;\;\;\;\; \text{all $i$ in bulk}\label{bulkL}
    \\
    \Lambda^{(+)}_0 &= \bar{\Lambda} \;\;\;\;\; i=0\label{0L}
    \\
    \Lambda^{(-)}_{N-1} &= \bar{\Lambda} \;\;\;\;\; i=N-1\label{N-1L} \; .
\end{align}

As mentioned earlier, one may rescale the coordinates to fix the value of exactly one of the conformal factors. In vacuum, then, with all $\bar{T}^{(i)\mu}_{\;\;\;\;\;\;\nu}=0$, the equations of motion (given a choice for the interaction coefficients and provided that $\bar{G}^\mu_{\;\nu}=-\bar{\Lambda}\delta^\mu_\nu$) specify a system of $N$ algebraic equations for $N$ variables: namely, $\bar{\Lambda}$, as well as the $N-1$ remaining unfixed conformal factors  (see Eq. \eqref{W_comps_prop}). One may in principle solve these equations to obtain the vacuum structure of the corresponding multi-metric theory (coordinate rescaling can always be used to fix the overall normalisation). In general, there may be multiple solutions, the physical ones being those where the conformal factors are real; the number of such physical solutions is in general dependent on both $N$ and $\Tcoeffs$.

 We further note that the splitting of the $\Lambda$'s given in Eqs. \eqref{bulkL}-\eqref{N-1L} means that, unless $\bar{\Lambda}=0$, or unless one includes a bare cosmological constant \emph{not} arising from the $W$-tensor on the boundary sites to account for the missing terms, it is impossible to find solutions where there is a constant ratio between the conformal factors throughout the chain of interacting metrics i.e. where $a_{i+1}/a_i=C \; \forall \, i$.

These proportional solutions contain multi-metric analogues of all of the known GR black hole solutions. In particular, in $D$-dimensions, if one takes any of the Myers-Perry metrics \cite{MP} as their $\bar{g}_{\mu\nu}$, the proportional ansatz Eq. \eqref{propBG} will be a solution of the multi-metric theory, provided that there exists a physical solution to the equations for the conformal factors. Within this general class of solutions, there are contained many special cases that are interesting to consider on their own. We shall look at a couple of concrete examples in $D=4$ for demonstrative purposes.

\subsubsection{Deconstructed black string in 4d}\label{Sec:BS}

As a straightforward example, but one with an interesting physical interpration, one may take $\bar{g}_{\mu\nu}$ to be the Schwarzschild metric. One then has the situation where the multi-gravity metrics are:
\begin{equation}\label{Sch}
    \extd s^2_{(i)} = a_i^2 \left[-\left(1-\frac{r_s}{r}\right)\extd\eta^2 + \frac{\extd r^2}{\left(1-\frac{r_s}{r}\right)} + r^2\extd\Omega_{(2)}^2\right] \; ,
\end{equation}
with $r_s$ the Schwarzschild radius and $\extd\Omega_{(2)}^2$ the line element for the $2$-sphere. This ansatz solves the vacuum equations, provided that the conformal factors are such that $\bar{\Lambda}=0$.

The physical meaning of the various conformal factors in this scenario is clear: suppose we make the coordinate changes $r\rightarrow \tilde{r}$ and $\eta\rightarrow t$ defined by $r=\tilde{r}/a_i$ and $\extd\eta=\extd t/a_i$, then the $i$-th metric becomes:
\begin{equation}
    \extd s^2_{(i)} = -\left(1-\frac{r_sa_i}{\tilde{r}}\right)\extd t^2 + \frac{\extd\tilde{r}^2}{\left(1-\frac{r_sa_i}{\tilde{r}}\right)} + \tilde{r}^2\extd\Omega_{(2)}^2 \; ,
\end{equation}
we see that an observer minimally coupled to the $i$-th metric would see a black hole whose Schwarzschild radius is scaled by $a_i$. In principle, we can imagine having separate observers minimally coupled to each metric, who would each report seeing a black hole with a different sized horizon according to the vacuum structure dictated by $\Wi=0$.

This situation is reminiscent of the well-known black string solutions in higher dimensional gravity, where Schwarzschild hypersurfaces are glued together to form an extra dimension (see \cite{Ricci-flat-branes} for the general $p$-brane solution). The multi-metric solution where the metrics are all conformally Schwarzschild is precisely the dimensional deconstruction \cite{deconstructing_dims,Deconstructing} of these black strings, where each metric is to be thought of as corresponding to a discrete location in the extra compact dimension (which must be an \emph{interval}, rather than an orbifold \cite{ClockworkCosmo}), and the information regarding the geometry of the extra dimension is encoded in the structure of the conformal factors. 

For example, one may choose to work with a \emph{clockwork} theory \cite{ClockworkGrav,Deconstructing,ClockworkCosmo}, which is special amongst the general multi-metric constructions in that it further imposes that the vacuum structure is such that the conformal factors possess a hierarchy, leading to one end of the chain of metrics being exponentially suppressed compared to the other i.e. something like:
\begin{equation}\label{VacStructure}
    a_i = \frac{a_0}{q^i} \, ,
\end{equation}
with $q\gtrsim1$. The idea is that by coupling matter to the suppressed end of the chain, we engineer a suppressed coupling to the surviving massless graviton, since one can show that the structure of the zero-mode is directly proportional to the vacuum structure \cite{ClockworkGrav,ClockworkCosmo} -- we will see this explicitly in Section \ref{Sec:linear} . This way, one can imagine a situation whereby the fundamental scale of the theory is small, but matter interactions with the massless graviton are still at the Planck scale, with a view to solving the hierarchy problem (see \cite{Rattazzi,Choi_relaxion_clockwork,Giudice,Disassembling_the_clockwork,Giudice_rebuttal} and references therein for an overview of this idea in non-gravitational contexts).

One must choose their $\beta$'s in the manner prescribed by Eq. \eqref{vac_eqs} to ensure that such a vacuum solution exists; two example models were constructed in \cite{ClockworkCosmo}, one being essentially a deconstruction of the 5D Randall-Sundrum (RS1) braneworld model \cite{RS1,RS2,Carsten_review,Langlois_review}. 

With this vacuum structure, we see that the system is solved by a series of 4D Schwarzschild metrics whose horizon sizes decrease by a factor $q$ as one moves along the chain of interacting metrics. This looks a lot like the AdS black string in 5D \cite{BW_BHs}, where the system is solved by a metric with Schwarzschild hypersurfaces multiplied by an exponential factor that decays smoothly as one moves along the extra dimension. Indeed, for the parameter choices corresponding to the deconstructed RS1 model considered in \cite{ClockworkCosmo}, the solution \eqref{Sch} is precisely the deconstruction of the RS black string. The general 5D continuum limit of 4D multi-gravity \cite{ClockworkCosmo,Deconstructing} is a more complicated scalar-tensor braneworld, but the idea remains the same, with the continuum theory admitting Ricci-flat hypersurfaces.

\subsubsection{Kerr-Newman-(Anti-)de Sitter black holes in 4d}\label{Sec:KNdS_prop}

The most general black hole solution one is able to write down in 4D GR, as a consequence of the various no-hair theorems (see e.g. \cite{no_hair_1,no_hair_2,no_hair_3}), is the Kerr-Newman-(A)dS metric, which describes a rotating, charged black hole living in a universe with non-zero cosmological constant. It is a solution to the Einstein-Maxwell equations, so for our multi-metric scenario we include the following matter action:
\begin{equation}
    S_M = \sum_{i=0}^{N-1}\int_{\mathcal{M}_D} F^{(i)}\wedge\hodge^{(i)}F^{(i)} \; ,
\end{equation}
where the $F^{(i)}=\extd A^{(i)}$ are separate electromagnetic field strengths on each site, given as exterior derivatives of the corresponding $U(1)$ connections $A^{(i)}$. In components, the field strengths are $F^{(i)}_{\mu\nu}=\partial_\mu A^{(i)}_\nu - \partial_\nu A^{(i)}_\mu$ and the corresponding energy-momentum tensors are:
\begin{equation}
    \Tmunui = F^{(i)\mu}_{\;\;\;\;\;\alpha}F_\nu^{(i)\alpha}-\frac14\delta^\mu_\nu F^{(i)}_{\alpha\beta}F^{(i)\alpha\beta} \; .
\end{equation}

The common metric $\bar{g}_{\mu\nu}$ is typically written in Boyer-Lindquist coordinates as \cite{BH_book,KNdS1,KNdS2}:
\begin{equation}\label{KNdS_4d}
\begin{split}
    \extd{\bar{s}}^2 &= -\frac{\Delta_r}{\rho^2\Xi^2}\left(\extd{t}-j\sin^2{\theta}\extd{\phi}\right)^2 + \frac{\rho^2}{\Delta_r}\extd{r}^2
    \\
    &+ \frac{\Delta_\theta}{\rho^2}\extd{\theta}^2 + \frac{\Delta_\theta \sin^2{\theta}}{\rho^2\Xi^2}\left(j\extd{t}-(r^2+j^2)\extd{\varphi}\right)^2 \; ,
\end{split}
\end{equation}
where $j$ is the rotation parameter\footnote{The rotation parameter is usually written $a$, but we choose $j$ to avoid confusion with the conformal factors.} and the various functions are defined as
\begin{align}
    \Delta_r &= \left(r^2+j^2\right)\left(1-\frac{\bar{\Lambda}}{3}r^2\right) - r_s r + r_Q^2 \; ,
    \\
    \Delta_\theta &= 1 + \frac{\bar{\Lambda}}{3}j^2\cos^2{\theta} \; ,\label{delta_th}
    \\
    \Xi &= 1 + \frac{\bar{\Lambda}}{3}j^2 \; ,
    \\
    \rho^2 &= r^2 + j^2\cos^2{\theta} \; ,\label{rho2}
\end{align}
with $r_Q$ a scale related to the electric charge in a manner that will be fixed below. If $\bar{\Lambda}>0$, the metric is Kerr-Newman-dS, while if $\bar{\Lambda}<0$ the metric is Kerr-Newman-AdS.

With this choice for $\bar{g}_{\mu\nu}$, the Einstein tensor acquires contributions from both the cosmological constant and the charge. The former is accounted for by the $W$-tensor in exactly the manner previously described, while the latter is supplied by the non-trivial electromagnetic fields:
\begin{equation}\label{EM field}
    A^{(i)} = \frac{Q_i r}{\rho^2\Xi}\left(\extd t + j\sin^2{\theta}\extd \varphi\right) \; ,
\end{equation}
provided that the charges are expressed in terms of $r_Q$ as
\begin{equation}
    Q_i = \sqrt{2}M_i a_i r_Q \; .
\end{equation}
One can check that the field equations for the electromagnetic fields, $\nabla^{(i)\mu}F_{\mu\nu}^{(i)}=0$, are also satisfied for this choice of $A_\mu^{(i)}$.

This solution contains all the other proportional black hole solutions of the $D=4$ multi-metric theory, in the various limits where one takes combinations of the characteristic parameters ($r_s, r_Q, \bar{\Lambda}, j$) to 0. For example, taking $r_Q\rightarrow 0$ gives us the Kerr-(A)dS solution, if we also take $j\rightarrow 0$ we get Schwarzschild-(A)dS, then taking $\bar{\Lambda}\rightarrow 0$ we recover the Schwarzschild (black string) solution from earlier. If instead we keep $r_Q$ but send $j$ and $\bar{\Lambda}$ to 0, we get Reissner-Nordstrom. Finally, keeping $\bar{\Lambda}$ but sending all relevant black hole parameters to 0 we recover the static coordinate form of the dS vacua that were previously constructed in FLRW coordinates in \cite{ClockworkCosmo}.

\subsection{Non-Proportional Solutions}\label{Sec:non prop sols}

The existence of the charged black hole solutions described above is reliant on one having $N$ non-interacting copies of the Maxwell action for entirely separate electromagnetic fields, each minimally coupled to its own separate metric, yet each taking the same special field configuration that cancels the Einstein tensor contribution given in Eq. \eqref{EM field}. As alluded to earlier, such a situation is not necessarily realistic, so we are motivated to try and find solutions where there is, say, a single matter sector coupled to only one distinguished metric (such a situation, again, is natural from the deconstruction perspective, where it is analogous to placing matter on a brane at some distinguished location in the extra dimension). 

In this scenario, the proportional ansatz breaks, so the solutions that we find describe situations where the various metrics cannot be simultaneously diagonalised. For bigravity ($N=2$), the corresponding \emph{non-bidiagonal} solutions were constructed in a series of works that established the exact set of metrics describing static, charged, rotating and asymptotically (A)dS black holes of this type in $D=4$ dimensions \cite{spherical_sym_sols,charged_BHs_bigravity,rotating_BHs_bigravity,Rotating_AdS_bigravity}. Here, we generalise the bigravity results to the multi-metric theory, establishing a wide class of \emph{non-multidiagonal} black hole solutions in arbitrary dimension. Unlike the proportional solutions from before, the field equations do not reduce simply to equivalent copies of the standard GR field equations; nevertheless, the metrics are still patterned sitewise as GR solutions, so we refer to these solutions as `GR-adjacent'. 

\subsubsection{Rotating (A)dS black holes in arbitrary dimension}\label{Sec:rot_nonprop}

We first look for solutions describing rotating but non-charged $D$-dimensional black holes that are asymptotically (A)dS. We do not include charge for the arbitrary $D$ case, as generally this changes the form of the metric non-trivially whenever there is also rotation, although we shall soon provide the  $D=4$ solution where it is simple enough to include both charge and rotation, as well as the charged but non-rotating solution in arbitrary $D$. It proves useful to work in the metric formalism here, and express the metrics in Kerr-Schild coordinates, where the line elements read \cite{Gibbons_KdS,Rotating_AdS_bigravity}:
\begin{equation}\label{KdS}
   \gi = a_i^2\left(\bar{g}_{\mu\nu} + 2\phi_i l_\mu l_\nu\right) \; .
\end{equation}
Here, $\bar{g}_{\mu\nu}$ is taken to be the metric of $D$-dimensional (A)dS space, $\phi_i$ are scalar functions whose form will be given shortly, and $l$ is a null vector that is tangent to a null-geodesic congruence on (A)dS. Following the conventions of \cite{Gibbons_KdS}, the (A)dS metric is best expressed in ellipsoidal coordinates, where the line element takes the following form:
\begin{equation}\label{AdS_ellipsoid}
    \begin{split}
        \dd\bar{s}^2 = &-W\left(1-\lambda r^2\right) \dd t^2 + F \dd r^2 
        \\
        &+ \sum_{k=1}^n \frac{r^2 + j_k^2}{\Xi_k}\left(\dd\mu_k^2 + \mu_k^2\dd\varphi_k^2\right) 
        \\
        &+ \frac{\lambda}{W\left(1-\lambda r^2\right)}\left[\sum_{k=1}^n\frac{(r^2+j_k^2)\mu_k\dd\mu_k}{\Xi_k}\right]^2 \; .
    \end{split}
\end{equation}
Some elaboration is required here: the coordinate system comprises a time coordinate $t$, a radial coordinate $r$, $\lfloor (D-1)/2\rfloor$ azimuthal coordinates $\varphi_k$, and $\lfloor D/2\rfloor$ coordinates $\mu_k$ that satisfy $\sum_{k=1}^{\lfloor D/2\rfloor}\mu_k^2=1$. The sums then run to $n=\lfloor D/2\rfloor$, although if $D$ is even then there is one fewer azimuthal coordinate relative to when $D$ is odd, so one should in the even dimensional case set $\varphi_n=\dd\varphi_n=0$. The $j_k$ are at this stage simply $\lfloor(D-1)/2\rfloor$ parameters that describe the ellipticity of the spacetime foliation ($j_k=0$ then just gives the (A)dS metric in spherical coordinates), although they will become genuine rotation parameters after extending the metrics by the null vector $l$. In the even dimensional case, one should also therefore set $j_n=0$. The various functions appearing in Eq. \eqref{AdS_ellipsoid} are:
\begin{align}
    \lambda &= \frac{2\Lambda}{(D-2)(D-1)} \; ,
    \\
    \Xi_k &= 1 + \lambda j_k^2 \; ,
    \\
    W &= \sum_{k=1}^n \frac{\mu_k^2}{\Xi_k} \; ,
    \\
    F &= \frac{r^2}{1-\lambda r^2}\sum_{k=1}^n \frac{\mu_k^2}{r^2+j_k^2} \; ,
\end{align}
and the required null vector (and its dual 1-form) that is tangent to a null-geodesic congruence in this spacetime is:
\begin{align}
    l^\mu\partial_\mu &= -\frac{1}{1-\lambda r^2} \partial_t + \partial_r -\sum_{k=1}^n \frac{j_k}{r^2+j_k^2}\partial_{\varphi_k}  \; ,
    \\
    l_\mu\dd x^\mu &= W\dd t + F\dd r -\sum_{k=1}^n \frac{j_k\mu_k^2}{\Xi_k}\dd\varphi_k \; .
\end{align}
Finally, the scalar functions $\phi_i$ are given by:
\begin{equation}\label{phis}
    \phi_i = \frac{r_{s,i}}{2U} \; ,
\end{equation}
with each metric now allowed its own independent Schwarzschild radius $r_{s,i}$, and where the function $U$ differs between the even and odd dimensional cases \cite{Rotating_AdS_bigravity,Gibbons_KdS}:
\begin{align}
    U &= \sum_{k=1}^{n} \frac{\mu_k^2}{r^2 + j_k^2} \prod_{s=1}^n \left(r^2 + j_s^2\right) \;\;\;\;\;\;\; \text{(D is odd)}
    \\
    U &= r\sum_{k=1}^{n} \frac{\mu_k^2}{r^2 + j_k^2} \prod_{s=1}^{n-1} \left(r^2 + j_s^2\right) \;\;\;\;\; \text{(D is even)}
\end{align}

In principle, one could similarly allow for $(i)$ labels on the rotation parameters and cosmological constants, though the field equations force these to be equal on all sites \cite{Rotating_AdS_bigravity}, else $M_i^{D-2}\Gi+\Wi=0$ leads to an inconsistency (unless one has all $j_k=0$, as we shall see).

With the ansatz Eq. \eqref{KdS} for the metrics, the Einstein tensors are simply:
\begin{equation}\label{Einstein tens}
    \Gi = -\frac{\Lambda}{a_i^2}\delta^\mu_\nu \; ,
\end{equation}
as we had for the proportional solutions. 

The utility of writing the metrics in Kerr-Schild form lies in how it simplifies the calculation of the $W$-tensors: the fact that $l$ is null with respect to both $\bar{g}_{\mu\nu}$ and $\gi$ (i.e. $l^\mu l_\mu=0$) means that its contribution to the interaction building block matrices $S_{i\rightarrow j}$ is nilpotent, leading to an early truncation in the expansion of the matrix square root. In particular, one may show that, with the metrics given by Eq. \eqref{KdS}, the $m$-th power of $S_{i\rightarrow j}$ is simply \cite{Rotating_AdS_bigravity}:
\begin{equation}
    (S_{i\rightarrow j}^m)^\mu_{\;\nu} = a_j^m a_i^{-m}\left[\delta^\mu_\nu - m\left(\phi_i-\phi_j\right)l^\mu l_\nu\right] \; .
\end{equation}
The next step is to substitute this into Eq. \eqref{Y_def} to obtain the form of the $Y_{(m)}$ matrices that enter the $W$-tensors. The null character of $l$ helps us here also, since it means that the trace of $S_{i\rightarrow j}$ picks up only the contribution from $\delta^\mu_\nu$, so the elementary symmetric polynomials are:
\begin{equation}
    e_n(S_{i\rightarrow j})=a_j^n a_i^{-n}\binom{D}{n} \; .
\end{equation}
Using this in Eq. \eqref{Y_def}, along with the binomial coefficient identities
\begin{align}
    \sum_{n=0}^m(-1)^n\binom{D}{n} &= (-1)^m\binom{D-1}{m} \; ,
    \\
    \sum_{n=0}^m n (-1)^n \binom{D}{n} &= D (-1)^m \binom{D-2}{m-1} \; ,
\end{align}
one finds that:
\begin{equation}\label{Kerr_Schild_Ys}
\begin{split}
    Y_{(m)\nu}^\mu(S_{i\rightarrow j}) &= (-1)^m a_j^m a_i^{-m}\\&\times \left[\binom{D-1}{m}\delta^\mu_\nu + \binom{D-2}{m-1}\left(\phi_i-\phi_j\right) l^\mu l_\nu \right] \; .
\end{split}
\end{equation}
One can then substitute into Eq. \eqref{metric eqs} to determine the components of the $W$-tensors. Since for chain type interactions the only building block matrices present in the potential are the nearest neighbour ones -- namely, $S_{i\rightarrow i+1}$ and $S_{i-1\rightarrow i}$, Eq. \eqref{metric eqs} involves only two sums:\pagebreak
\begin{align}
    \Wi &= \sum_{m=0}^D(-1)^m\beta_m^{(i,i+1)}Y_{(m)\nu}^\mu(S_{i\rightarrow i+1})\nonumber
    \\
    &+\sum_{m=0}^D (-1)^m\beta_{D-m}^{(i-1,i)}Y_{(m)\nu}^\mu (S_{i\rightarrow i-1}) \; ,\label{metric W}
\end{align}
recalling from section \ref{Sec:Review} that $S_{i-1\rightarrow i}^{-1}=S_{i\rightarrow i-1}$. With the $Y_{(m)}$ given by Eq. \eqref{Kerr_Schild_Ys}, and the $\phi_i$ given by Eq. \eqref{phis}, the explicit expression for the components becomes:
\begin{widetext}
\begin{equation}\label{W_nondiag}
    \begin{split}
        \Wi &= \delta^\mu_\nu \left[\sum_{m=0}^D\beta_m^{(i,i+1)}\binom{D-1}{m}a_{i+1}^m a_i^{-m} + \sum_{m=0}^D\beta_{D-m}^{(i-1,i)}\binom{D-1}{m}a_{i-1}^m a_i^{-m} \right]
        \\
        &+ \frac{l^\mu l_\nu}{2U} \left[\frac{a_{i+1}}{a_i}\left(r_{s,i}-r_{s,i+1}\right)\Sigma_i^{(+)} + \frac{a_{i-1}}{a_i}\left(r_{s,i}-r_{s,i-1}\right)\Sigma_i^{(-)}\right] \; ,
    \end{split}
\end{equation}
\end{widetext}
where we have defined:
\begin{align}
    \Sigma^{(+)}_{i} &=  \sum_{m=0}^D \beta_m^{(i,i+1)}\binom{D-2}{m-1}a_{i+1}^{m-1}a_i^{1-m} \; ,\label{sig_p}
    \\
    \Sigma^{(-)}_{i} &= \sum_{m=0}^D \beta_{D-m}^{(i-1,i)}\binom{D-2}{m-1}a_{i-1}^{m-1}a_i^{1-m} \; .\label{sig_m}
\end{align}

The part of the $W$-tensor proportional to $\delta^\mu_\nu$ is exactly as in Eq. \eqref{W_comps_prop} for the proportional solutions, but now there are additional off-diagonal terms proportional to $l^\mu l_\nu$. Clearly, for the ansatz Eq. \eqref{KdS} to be a solution to the multi-metric vacuum equations, since the Einstein tensor is diagonal, we need these off-diagonal $W$-tensor components to vanish. There are a few ways of achieving this, and we shall go through them in order of increasing complexity. 

The simplest is to take all of the Schwarzschild radii to be the same,
\begin{equation}\label{same_rs}
    r_{s,i}=r_s \;\;\; \forall \, i \; ,
\end{equation}
which just recovers the proportional solution from before. 

The second is to make all of the $\Sigma_i$'s vanish:
\begin{align}
    \Sigma_i^{(+)} &= 0 \; ,\label{nondiag1}
    \\
    \Sigma_i^{(-)} &= 0 \;\;\; \forall \, i \; .\label{nondiag2}
\end{align}
These conditions are polynomial equations that fix the ratios of neighbouring conformal factors. Furthermore, due to the recurrence relation for the binomial coefficients:
\begin{equation}
    \binom{n}{k} = \binom{n-1}{k} + \binom{n-1}{k-1} \; ,
\end{equation}
they cause part of the sum defining the diagonal components of the $W$-tensors in Eq. \eqref{W_nondiag} to vanish also. Thus, the conditions \eqref{nondiag1} and \eqref{nondiag2} strengthen Eqs. \eqref{CCbars}, which give the value of the cosmological constant, so as to include only the non-vanishing parts of the diagonal $W$-tensor components:
\begin{equation}\label{Reduced_CC}
    \begin{split}
        &\Bigg[\sum_{m=0}^{D} \beta_m^{(i,i+1)} \binom{D-2}{m}a_{i+1}^m a_{i}^{-m} \\  &+  \sum_{m=0}^{D} \beta_{D-m}^{(i-1,i)} \binom{D-2}{m} a_{i-1}^m a_{i}^{-m} \Bigg] = \frac{\Lambda M_i^{D-2}}{a_i^2} \; .
    \end{split}
\end{equation}
For a solution to exist, the same ratios of conformal factors that solve Eqs. \eqref{nondiag1} and \eqref{nondiag2} must \emph{also} satisfy Eq. \eqref{Reduced_CC}, which can only happen when the $\beta_m^{(i,i+1)}$ are finely tuned to allow for this possibility. That is to say, Eq. \eqref{Reduced_CC} acts as a constraint on which multi-metric theories permit this class of solutions. If the parameters are chosen such that this constraint is satisfied, then one has succeeded in constructing their non-multidiagonal black hole. For example, if one wishes to find a solution with $\Lambda=0$, one only needs to fix $\beta_0$ and $\beta_D$ in terms of the other $\beta$'s and conformal factor ratios to satisfy the constraint \eqref{Reduced_CC}.

In bigravity, these are the only two available options. We see this by realising that the sum involving $\beta_{D-m}$ in the $(i+1)$-th off-diagonal $W$-tensor is proportional to the sum involving $\beta_m$ in the $i$-th off-diagonal $W$-tensor. Explicitly, one has that:
\begin{equation}\label{Sig_pm}
     \Sigma^{(-)}_{i+1} = \left(\frac{a_{i}}{a_{i+1}}\right)^{D-2} \Sigma^{(+)}_{i} \; ,
\end{equation}
so when $N=2$ and the only terms present are $\Sigma^{(+)}_{0}$ and $ \Sigma^{(-)}_{1}$ (since $\beta_m^{(-1,0)}=\beta_m^{(N-1,N)}=0$), one sum vanishing implies the other does too. Indeed, our equations recover exactly the bigravity results in e.g. \cite{rotating_BHs_bigravity,Rotating_AdS_bigravity} when one takes $N=2, D=4$ and $\Lambda=0$\footnote{In $D=4$, as we will shortly see, the coordinates of Eq. \eqref{AdS_ellipsoid} can be parametrised by $\mu_1=\sin\theta$ and $\mu_2=\cos\theta$, leading to a pair of metrics that are precisely those in \cite{Rotating_AdS_bigravity}, and are a coordinate transformation away from those written in Eddington-Finkelstein coordinates in \cite{rotating_BHs_bigravity}.}. 

In the multi-metric scenario, however, whenever one has $N>2$, there are more $\Sigma^{(+)}_{i}$ to play with, and these do not necessarily all have to be 0 \emph{even if} some of the others are. More precisely, if one has vanishing $\Sigma^{(+)}_{I-1}$ for some specific $i=I$, this implies only the vanishing of $\Sigma^{(-)}_{I}$ and nothing else, so the requirement that the off-diagonal parts of $W^{(I)\mu}_{\;\;\;\;\;\;\;\nu}$ vanish may still be satisfied by \emph{either} $r_{s,I}=r_{s,I+1}$ or $\Sigma^{(+)}_{I}=0$ (which is independent of $\Sigma^{(+)}_{I-1}$). Therefore, in the multi-metric theory with $N>2$, the most general way to solve the vacuum equations is to allow for combinations of both $\Sigma^{(+)}_{k}=0$ and $r_{s,k}=r_{s,k+1}$, for different $k\subset i$. This corresponds to the situation where some -- but not all -- of the metrics can be simultaneously diagonalised. Those sites where $\Sigma^{(+)}_{k}=0$ then each fix $a_{k+1}/a_k$ only, and feed in to modify only the $k$-th and $(k+1)$-th equations for the cosmological constant in the manner described by Eq. \eqref{Reduced_CC}, while the rest of those equations are unchanged i.e. still involve $\binom{D-1}{m}$ instead of $\binom{D-2}{m}$. If there are $n$ total sites that have $\Sigma_k^{(+)}=0$ ($n$ can therefore be at most $N-1$), $n+1$ of the conformal factors are a priori fixed before checking whether these cosmological constant equations are satsified (the additional 1 is fixed by rescaling the coordinates). Then, the $N$ diagonal equations split into $N-n$ algebraic equations for $\Lambda$ and the remaining $N-n-1$ free conformal factors, as well as $n$ equations that become constraints on the $\beta_m^{(i,i+1)}$ parameters of the theory. Again, this means that only for finely tuned parameters can these solutions exist. Indeed, the solutions form a set of measure zero in the $\beta_m^{(i,i+1)}$ parameter space.

This third and most general branch of solutions (which we dub the ``partially proportional'' branch) is fiddly and awkward to deal with; in practice, we shall stick to considering the first two branches of solutions i.e. where either all the Schwarzschild radii are the same (the proportional solutions) or all the $\Sigma_i$'s vanish (the non-proportional solutions). We note, however, that a subtlety arises on the overlap of these first two branches, when one has \emph{both} $r_{s,i}=r_{s,i+1}$ and $\Sigma^{(+)}_{i}=0$ for every $i$. In this scenario, one cannot accept the solution as valid: as we will demonstrate in section \ref{Sec:linear}, the masses of all the linearised perturbations vanish, and as mentioned in section \ref{Sec:Intro}, theories involving multiple interacting \emph{massless} gravitons are known to be pathological \cite{no_interacting_massless_gravitons}.

\subsubsection{4d Kerr-Newman-(Anti-)de Sitter revisited}\label{Sec:KNdS_nonprop}

In $D=4$, as we mentioned, it is not difficult to extend the above analysis to include charge, providing a non-proportional generalisation of the Kerr-Newman-(A)dS metric from section \ref{Sec:KNdS_prop}. In the $D=4$ case of Eq. \eqref{AdS_ellipsoid}, there is only one azimuthal coordinate $\varphi_1=\varphi$, only one rotation parameter $j_1=j$ (hence one $\Xi_1=\Xi$), and only two $\mu_k$ coordinates, which we can parametrise without loss of generality as \cite{Gibbons_KdS}:
\begin{equation}
    \mu_1=\sin\theta \; , \;\;\; \mu_2 = \cos\theta \; .
\end{equation}
With these definitions, the (A)dS metric explicitly takes the form:
\begin{equation}
    \begin{split}
        \dd\bar{s}^2 = &-\frac{\left(1-\lambda r^2\right)\Delta_\theta \dd t^2}{\Xi} + \frac{\rho^2\dd r^2}{\left(1-\lambda r^2\right)\left(r^2+j^2\right)}
        \\
        &+ \frac{\rho^2\dd\theta^2}{\Delta_\theta} + \frac{\left(r^2+j^2\right)\sin^2\theta\dd\varphi^2}{\Xi} \; ,
    \end{split}
\end{equation}
where $\Delta_\theta$ and $\rho^2$ are as in Eqs. \eqref{delta_th} and \eqref{rho2}. The null vector and 1-form are given by:
\begin{align}
    l^\mu\partial_\mu &= -\frac{1}{1-\lambda r^2} \partial_t + \partial_r - \frac{j}{r^2+j^2}\partial_{\varphi}  \; ,
    \\
    l_\mu\dd x^\mu &= \frac{\Delta_\theta\dd t}{\Xi} + \frac{\rho^2\dd r}{\left(1-\lambda r^2\right)\left(r^2+j^2\right)} -\frac{j\sin^2\theta\dd\varphi}{\Xi} \; ,
\end{align}
while the function $U$ becomes:
\begin{equation}
    U=\frac{\rho^2}{r} \; .
\end{equation}

As in section \ref{Sec:KNdS_prop}, to incorporate charge we must include copies of the Maxwell action for the matter sector,
\begin{equation}
    S_M = \sum_{k\subset i}\int_{\mathcal{M}_D} F^{(k)}\wedge\hodge^{(k)}F^{(k)} \; ,
\end{equation}
only now we allow the freedom to couple to only some subset of sites $k\subset i$, rather than to all of them. This of course breaks the proportional ansatz, but we are looking for the non-/partially proportional solutions anyway. 

In $D=4$, the only effect of the charge is to modify the scalar functions $\phi_i$ to \cite{Rotating_AdS_bigravity}:
\begin{equation}\label{charged_phi}
    \phi_i = \frac{1}{2U}\left(r_{s,i}-\frac{r_{Q,i}}{r}\right) \; ,
\end{equation}
where each metric is now allowed an independent $r_{Q,i}$ as well as $r_{s,i}$. The coordinate transformation that retrieves the Boyer-Lindquist form of the metric, Eq. \eqref{KNdS_4d}, from the Kerr-Schild form is given in \cite{Gibbons_KdS}. We note that the extension of $\phi_i$ to include charge while keeping rotation is particularly simple in $D=4$; this is assuredly not the case in higher dimensions, where even in standard GR the analogue of the Kerr-Newman metric for $D>4$ is not known, so there is no safe starting point to begin to look for the corresponding multi-metric solutions. This is why we treat the charged and rotating cases separately outside of $D=4$. 

To account for the additional contributions to the Einstein tensors on the charged sites, the corresponding electromagnetic fields must take the form:
\begin{equation}
    A^{(k)} = \frac{Q_k r}{\rho^2} l_\mu\dd x^\mu \; ,
\end{equation}
where as before
\begin{equation}
    Q_k = \sqrt{2}M_k a_k r_{Q,k} \; .
\end{equation}
Of course, only the sites $k\subset i$ that have an electromagnetic field coupling have non-vanishing $r_{Q,i}$, so the multi-metric system comprises Kerr-Newman-(A)dS metrics on those sites with charge, and Kerr-(A)dS metrics on those without.

Since the only alteration wrought by the inclusion of charge in the system is the new form of $\phi_i$ given by Eq. \eqref{charged_phi}, the only changes to the $W$-tensor components are in the part multiplying $l^\mu l_\nu$, where $r_{s,i}$ is shifted to $r_{s,i}-r_{Q,i}/r$. Therefore, in line with the discussion of the previous subsection, there are again in principle (depending on the parameters of the theory) three classes of solutions, corresponding to the three ways in which one can make these off-diagonal components vanish. The proportional solutions have the same $r_s$ \emph{and} $r_Q$ on every site, the non-proportional solutions have $\Sigma^{(+)}_i=0$ on every site, and the partially proportional solutions have a combination of the two for different $k\subset i$. These three classes of Kerr-Newman-(A)dS solutions are then the most general GR-adjacent black holes in the $D=4$ multi-metric theory. 

We note that the proportional solutions only exist in this case when one has a separate matter sector coupled to each metric in the chain, which we already argued is unrealistic. If one wishes to solve the system with, say, a single copy of the Maxwell action coupled to one metric only, then the solution must lie in either the non-proportional or partially proportional branch, as $r_{Q,i}-r_{Q,i+1}$ cannot vanish when only one $r_Q$ is present. Therefore, charged and rotating black holes of this type \emph{only} exist in the $D=4$ multi-metric theory if the interaction coefficients are finely tuned to allow for this possibility.

\subsubsection{Charged (A)dS black holes in arbitrary dimension}\label{Sec:RN_nonprop}

Finally, we look for the $D$-dimensional solution for an asymptotically (A)dS, charged, but \emph{non}-rotating black hole. As alluded to earlier, when there is no rotation, one may in principle allow for different cosmological constants on each site, and so replace $\lambda\rightarrow\lambda_i$ in Eq. \eqref{AdS_ellipsoid}. However, this then spoils the utility of the Kerr-Schild ansatz, since the null vectors pick up their own $(i)$ indices, which means that the expressions for the building block matrices $S_{i\rightarrow j}$ no longer simplify in the manner they did before. To proceed, we must change tack; once again, the vielbein formalism comes in handy.

Following \cite{charged_BHs_bigravity}, it proves useful to now express the metrics in (advanced) Eddington-Finkelstein coordinates, where the line elements read:
\begin{align}\label{charged_metric}
        \dd s^2_{(i)} = a_i^2\Bigg[&-\left(1-\lambda_i r^2 -\frac{r_{s,i}}{r^{D-3}}+\frac{r_{Q,i}^2}{r^{2(D-3)}}\right)\dd v^2 \nonumber
        \\
        &+ 2\dd v\dd r + r^2\left(\dd\theta^2+\sin^2\theta\dd\varphi^2\right) \nonumber
        \\
        &+ r^2\cos^2\theta\dd\Omega^2_{(D-4)}\Bigg] \; ,
\end{align}
with $\dd\Omega_{(D-4)}^2$ the unit round metric on the $(D-4)$-sphere, given by:

\begin{equation}
    \dd\Omega_{(D-4)}^2 = \dd\psi_1^2 + \sum_{k=2}^{D-4}\left(\prod_{m=1}^{k-1}\sin^2\psi_m\right)\dd\psi_k^2 \; ,
\end{equation}

As before, to account for the extra contribution to the Einstein tensors on the sites with non-vanishing $r_{Q,i}$, the corresponding electromagnetic fields must take the profile:
\begin{equation}
    A^{(k)} = \sqrt{\frac{D-2}{2(D-3)}}\frac{Q_k}{r^{D-3}}\dd t \; ,
\end{equation}
again with:
\begin{equation}
   Q_k = \sqrt{2}M_k a_k r_{Q,k} \; .
\end{equation}

With the ansatz Eq. \eqref{charged_metric} for the metric, the tetrads have the following form:
\begin{align}
        e^{(i)0} &= \dd r -\frac{1}{2}\left(2-\lambda_i r^2 - \frac{r_{s,i}}{r^{D-3}}+\frac{r_{Q,i}^2}{r^{2(D-3)}}\right)\dd v
    \\
        e^{(i)1} &= \dd r +\frac{1}{2}\left(\lambda_i r^2 + \frac{r_{s,i}}{r^{D-3}}-\frac{r_{Q,i}^2}{r^{2(D-3)}}\right)\dd v
    \\
    e^{(i)2} &= r \dd\theta,
    \\
    e^{(i)3} &= r\sin\theta\dd\varphi,
    \\
    e^{(i)4} &= r\cos\theta \dd\psi_1,
    \\
    e^{(i)5} &= r\cos\theta\sin\psi_1 \dd\psi_2,
    \\
    \vdots\nonumber
\end{align}
which one can use to determine the components of the $W$-tensors by substituting into Eq. \eqref{W tensor}. 

The diagonal part of the $i$-th $W$-tensor is precisely as in Eq. \eqref{W_comps_prop} and again gives ($M_i^{D-2}/a_i^2$ times) the $i$-th effective cosmological constant $\Lambda_i$, in the usual manner. The only non-vanishing off-diagonal term is the $rv$ component, which becomes:
\begin{widetext}
    \begin{equation}\label{Wrv_charged}
        \begin{split}
             W^{(i)r}_{\;\;\;\;\;\;v} = \frac{1}{2}&\Bigg\{\frac{a_{i+1}}{a_i}\left[\frac{r_{s,i}-r_{s,i+1}}{r^{D-3}}+(\lambda_i-\lambda_{i+1})r^2 - \frac{r_{Q,i}-r_{Q,i+1}}{r^{2(D-3)}}\right] \Sigma_i^{(+)}
             \\
             &+ \frac{a_{i-1}}{a_i}\left[\frac{r_{s,i}-r_{s,i-1}}{r^{D-3}}+(\lambda_i-\lambda_{i-1})r^2 - \frac{r_{Q,i}-r_{Q,i-1}}{r^{2(D-3)}}\right] \Sigma_i^{(-)} \Bigg\} \; .
        \end{split}
    \end{equation}
\end{widetext}
Remarkably, both the charge and cosmological constant contributions factor nicely into the collection of terms multiplying $\Sigma_i^{(\pm)}$. The solutions then split into the usual three branches: the proportional solutions are those where one has $r_{s,i}=r_s \, , \; \Lambda_i=\Lambda$ and $r_{Q,i}=r_Q \; \forall\, i$, the non-proportional solutions are those with $\Sigma_i^{(+)}=0 \; \forall\, i$, and the partially proportional solutions are those with combinations of either on different sites.

For the non-proportional case, all of the effective cosmological constants are fixed by the diagonal part of the vacuum equations once $\Sigma_i^{(+)}=0$ has fixed the ratios of all the adjacent conformal factors. However, because the $\Lambda_i$ no longer have to be the same, the $\beta_m^{(i,i+1)}$ are now \emph{unconstrained}. 

For the partially proportional case, some, but not all, of the adjacent sites \emph{do} still need to have the same $\Lambda$, which, as previously, makes things more fiddly. The situation is as follows: if there are $n$ total sites $k\subset i$ which have $\Sigma_k^{(+)}=0$, then $n+1$ conformal factors are a priori fixed before considering the diagonal part of the vacuum equations. Furthermore, there are only $n+1$ independent $\Lambda_i$, as the sites that do not have $\Sigma_k^{(+)}=0$ must have $\Lambda_k=\Lambda_{k+1}$. Then, from the $N$ total diagonal equations, $n$ of them fix $n$ of the independent $\Lambda_i$, leaving only one free, so that one is left with $N-n$ algebraic equations for the remaining $N-n-1$ free conformal factors and remaining one free $\Lambda$. Again, the $\beta_m^{(i,i+1)}$ are unconstrained here, as the $n$ equations that constrained them in the rotating case now act instead to fix the initially free $\Lambda_i$.

Because of the lack of constraints on which parameters one must choose for these non-rotating solutions to exist, one can imagine being able to choose their model in such a way as to engineer essentially whatever effective cosmological constants one would like. In particular, taking the limit where $r_{s,i}$ and $r_{Q,i}$ go to 0, one is faced with the tantalising prospect of finding a multi-metric theory with a non-proportional dS vacuum whose effective cosmological constant on the `physical' metric i.e. the one matter couples to, is small but non-vanishing, with a view to addressing the cosmological constant problem. However, one expects that in order to do this, the present fine-tuning problem regarding the size of the cosmological constant would simply be transferred to a fine-tuning problem regarding the potential coefficients and/or vacuum structure, so nothing is actually alleviated. Furthermore, as we have seen, one requires the effective cosmological constants on each metric to be the same anyway, if one wishes the theory to admit rotating black holes, which likely comprise the majority of real, physical black holes that exist in nature. The argument is also a purely classical one; there is nothing to be said about quelling the contribution of the QFT vacuum energy. Therefore, it is unlikely that the cosmological constant problem may be addressed in this manner.

\subsection{Other Solutions and Remarks on Stability}\label{Sec:Hairy}

As we mentioned previously, the wide class of analytic black hole solutions constructed above were all built starting from metrics that are known to GR. Even the non- and partially proportional solutions that do differ from GR -- in the sense that the field equations are not simply $N$ copies of the standard GR equations, as is the case for the proportional solutions -- are patterned sitewise as GR solutions. However, in dRGT massive gravity and bigravity there also exist additional non-GR-adjacent black hole solutions, which one would expect to carry over to the multi-metric theory as well.

Firstly, however, there is one remaining GR-adjacent solution we missed that is worth a mention. Although we did not explicitly write down this solution, the BTZ black hole in $D=3$ \cite{BTZ} does admit multi-metric analogues in each of the proportional, non-proportional and partially proportional branches. This is most easily seen by expressing the BTZ metric in Kerr-(A)dS like coordinates (see e.g. \cite{BTZ_KdS_coords}) and following through exactly as before, the only changes resulting from the fact that there is no longer a $\theta$ coordinate. If one wishes to include charge in $D=3$, the electromagnetic field must also take a logarithmic profile, required to supply the Einstein tensor charge contribution \cite{ChargedBTZ}.

As for the non-GR-adjacent solutions, owing to the complexity of the field equations, solutions must generically be found numerically\footnote{Actually, there is a further class of analytic black hole solutions in $D$-dimensional dRGT massive gravity, for a particular theory whose non-dynamical reference metric takes a special degenerate form, and where only the first 4 interaction terms are included \cite{Cai,PV_crit,Ghosh} (see also \cite{holo_trans_sym,Adams,deg_ref}). However, when both metrics (or more) are dynamical, the degeneracy is of course problematic, so we shall leave these solutions alone (they are also not completely general i.e. do not include the full set of ghost free interactions when $D>4$).}. An important such class of numerically determined solutions, at least in bigravity, describing a family of asymptotically AdS black holes endowed with a cloud of massive graviton hair was found in \cite{Hairy_BH_AdS}, in a comprehensive analysis that also studied (for example) the case where the two metrics are diagonal but not proportional. This result was extended to asymptotically \emph{flat} hairy black holes in \cite{Hairy_BHs_flat}. The solutions are found by considering a generic static, spherically symmetric ansatz for the metrics and utilising the Bianchi constraint, Eq. \eqref{W constraint}, to reduce the field equations to a set of coupled first order ODEs that are then numerically integrated to determine the appropriate metric functions. A detailed derivation of these equations and their final form is publicly available in a Mathematica notebook online \cite{Hairy_website}. Again, one expects that these solutions should extend to the multi-metric theory, just with the precise form of the ODEs altered by the extra interactions, though we leave this calculation to future work. 

The solutions we constructed earlier, together with the proposed hairy solutions, comprise the full set of currently known black hole solutions of multi-metric gravity. It is natural to ask of these solutions the question of their stability: which (if any) of them are stable to perturbations, and if so, in what regimes of parameter space are they stable? 

Again, significant progress in this direction has already been made in both $D=4$ dRGT massive gravity and bigravity. There, it is known that the bidiagonal Schwarzschild(-dS) solution i.e. where both metrics are proportionally Schwarzschild(-dS), suffers from a spherically symmetric radial instability for certain values of the graviton mass \cite{GL_instability_bigravity,Kerr_instability_bigravity,GL_unified}. The instability takes precisely the same form as the Gregory-Laflamme (GL) instability that notoriously plagues black string solutions in higher dimensions \cite{GL_instability,Charged_GL_instability,AdS_GL_instability}. This should be unsurprising -- as we saw in section \ref{Sec:BS}, the proportional Schwarzschild solutions are just dimensionally deconstructed black strings. The bidiagonal Kerr solution also possesses this radial instability \cite{Kerr_instability_bigravity}, as well as a superradiant instability for the azimuthal modes, again dependent on the graviton mass \cite{Kerr_instability_bigravity,superradiance_ULS2,superradiance_massive_spin2,backreaction}. 

The non-bidiagonal Schwarzschild solution, on the other hand, seems to be linearly stable to metric perturbations \cite{GL_unified,Stability_nonbidiag}, although the perturbations take an unconventional, non-Fierz-Pauli form, and as of yet no analysis exists as to whether these solutions may contain (non-BD) ghosts. It is interesting to note that the particular combination of the interaction coefficients, $\Sigma_0^{(+)}=0$, that gives rise to the non-bidiagonal black hole solutions also shows up when one considers the theory's cosmological implications: namely, it is one of two possible ways to satisfy the Bianchi constraint when one tries to find FLRW solutions in dRGT/bigravity \cite{FRW_cosmology_dRGT,self_accelerating_branch_1,self_accelerating_branch_2} (the multi-metric extension is in \cite{ClockworkCosmo}, which further admits a partially proportional branch). The non-proportional branch of cosmological solutions (in dRGT/bigravity) is littered with pathologies e.g. not all massive graviton degrees of freedom propagate at linear level, ghosts appear at non-linear level etc. \cite{bigravity_review,mg_cosmo_perts,perts_open_FRW,nonlinear_instability_FRW,nonlinear_cosmo_stability,general_mass,viable_cosmo_bigravity}. However, these pathologies in cosmology appear to be intimately related to the symmetries of the FLRW background, which is potentially why they do not show up for the non-proportional black hole solutions. Also, the scale factors in the cosmological solutions are time dependent, whereas for our black hole solutions they are constant, so it is unclear whether one may even identify the $\Sigma_i^{(+)}=0$ branches in either scenario anyway.

One would expect that the dRGT/bigravity results for the non-proportional black hole solutions extend naturally to the full multi-metric theory, with the perturbations taking the same generic form as in \cite{GL_unified,Stability_nonbidiag}\footnote{Precisely, we expect that (e.g.) Eq. (11) in \cite{GL_unified} has instead of the term written $\mathcal{A}(r_S-r_f)$ a term that involves both $(r_{s,i}-r_{s,i+1})$ and $(r_{s,i}-r_{s,i-1})$, accounting for the interactions going in both directions.}. Therefore, the non-proportional multi-metric solutions should still be linearly stable. Determining the stability of the partially proportional branch may be more complicated, as on some sites the perturbations will acquire a standard Fierz-Pauli mass term while on others they will not. Due to the inherent complexity of these calculations, and the fact that, as we found, the interaction coefficients generally must be finely tuned to permit these branches of solutions anyway, an explicit determination of the stability of the non- and partially proportional multi-metric black hole solutions is left to future work.

This leaves us with the task of extending the results regarding the stability of the proportional solutions in bigravity to the general multi-metric scenario, which we will eventually get to in section \ref{Sec:stability}. To begin to address this point, however, we must first determine the dynamics of the graviton mass modes, which requires that we consider linear perturbations around the proportional backgrounds.

\section{Linearised perturbations and mass modes}\label{Sec:linear}

The notion of mass arises naturally in Minkowski space as a Casimir invariant of the Poincar\'e group relating to spacetime translations. In more general spacetimes with fewer symmetries, it is not always obvious how this notion should generalise. However, around the proportional solutions only, the spin-2 fluctuations all acquire a standard Fierz-Pauli mass term, and so give rise to a well-defined mass spectrum.

The full details of the linearisation procedure around a generic proportional solution are laid out in appendix \ref{App:linearisation}, following the formalism developed in \cite{consistent_spin2,Bigravity_gen_dim} for bigravity (it proves simpler to work in the metric formalism here). The metrics are expanded as:
\begin{align}\label{expanded metric}
    g^{(i)}_{\mu\nu} &= a_i^2\bar{g}_{\mu\nu} + a_i M_{i}^{-\frac{D-2}{2}} h^{(i)}_{\mu\nu} \; ,
\end{align}
where the normalisation is to ensure the kinetic terms are canonical. The resulting equations of motion for the perturbations read:
\begin{equation}\label{linearised eqs}
    \begin{split}
        \bar{\mathcal{E}}^{\alpha\beta}_{\;\mu\nu}h^{(i)}_{\alpha\beta} + \bar{\Lambda}h_{\mu\nu}^{(i)}
        +\frac{\mathcal{M}^2_{ij}}{2}&\left(h_{\mu\nu}^{(j)}-\bar{g}_{\mu\nu}h^{(j)}\right) \\ &= a_i M^{-\frac{D-2}{2}}_i\bar{T}^{(i)}_{\mu\nu} \; ,
    \end{split}
\end{equation}
where the Lichnerowicz operator is given by \cite{GL_instability_bigravity}:
\begin{equation}\label{Lichnerowicz}
\begin{split}
    \bar{\mathcal{E}}^{\alpha\beta}_{\;\mu\nu}h_{\alpha\beta} = \frac12\bigl[ &-\bar{\Box} h_{\mu\nu} + \bcovd[\mu]{\bcovd[\alpha]{h^\alpha_{\;\nu}}} + \bcovd[\nu]{\bcovd[\alpha]{h^\alpha_{\;\mu}}} \\ &- \bcovd[\mu]{\bcovd[\nu]{h}}
    + \bar{g}_{\mu\nu}\bar{\Box}h - \bar{g}_{\mu\nu}\bcovd[\alpha]{\bcovd[\beta]{h^{\alpha\beta}}} \\ &- 2\bar{R}^{\alpha\;\beta}_{\;\mu\;\nu}h_{\alpha\beta}\bigr] \; ,
\end{split}
\end{equation}
and all quantities have been rescaled by the appropriate powers of $a_i$, as in section \ref{Sec:prop sols}, so that all tensors in Eq. \eqref{linearised eqs} behave as if they live in the common background of $\bar{g}_{\mu\nu}$. For chain type interactions, the mass matrix, $\mathcal{M}^2$, for the rescaled perturbations in this common background, is tridiagonal, with the following non-vanishing components:
\begin{align}
    \mathcal{M}_{ii}^2 &= \frac{a_i^2}{M_i^{D-2}} \left(\Sigma_i^{(+)}\frac{a_{i+1}}{a_i} + \Sigma_i^{(-)}\frac{a_{i-1}}{a_i}\right) \label{Mii} \; ,
    \\
    \mathcal{M}_{i+1,i}^2&=\left(\frac{a_{i+1}}{a_i}\right)^{4-D}\mathcal{M}_{i,i+1}^2 = -\frac{a_i^2 \Sigma_i^{(+)}}{\left(M_{i+1}M_i\right)^{\frac{D-2}{2}}}\label{Mi1} \; .
\end{align}

As per the discussion in Section \ref{Sec:Review}, the mass eigenstates are 
linear combinations of the $h^{(i)}_{\mu\nu}$. To obtain them, we rotate to the field basis $\mathbf{H}_{\mu\nu}=(H^{(0)}_{\mu\nu},\hdots,H^{(N-1)}_{\mu\nu})$ in which the mass matrix is diagonalised, related to the original basis $\mathbf{h}_{\mu\nu}=(h^{(0)}_{\mu\nu},\hdots,h^{(N-1)}_{\mu\nu})$ via some $N\times N$ orthogonal matrix $O$ whose columns are the mass eigenvectors; that is,
\begin{align}
    \mathbf{h}_{\mu\nu} &= O \mathbf{H}_{\mu\nu}\label{mass modes} \;,\\
    O^T \mathcal{M}^2 O &= \text{diag}(m^2_0,\hdots,m^2_{N-1}) \; .
\end{align}

The structure of the potential means that the mass matrix will always possess one zero eigenvalue, whose associated eigenvector is proportional to the vacuum structure i.e. $O^{i0}\propto a_i$ \cite{prop_bg_multigrav,ClockworkGrav,ClockworkCosmo}. Generally, it is not possible (save for an exceptional case that we will consider shortly) to determine the higher mass eigenvalues/eigenvectors analytically \cite{Losonczi,Fonesca_tridiags}, although the expressions \eqref{Mii} and \eqref{Mi1} for the components allow one to readily determine them numerically on a case-by-case basis.

Regardless, by substituting Eq. \eqref{mass modes} into the linearised Einstein equations \eqref{linearised eqs} and multiplying by $O^T$ from the left, one finds the evolution equations for the mass modes:
\begin{equation}
\begin{split}
    \bar{\mathcal{E}}^{\alpha\beta}_{\;\mu\nu}H^{(i)}_{\alpha\beta} + \bar{\Lambda}H_{\mu\nu}^{(i)} + \frac{m^2_i}{2}&\left(H^{(i)}_{\mu\nu} - \bar{g}_{\mu\nu} H^{(i)}\right) \\ &= O^{ji}a_jM_{j}^{-\frac{D-2}{2}} \bar{T}^{(j)}_{\mu\nu} \; .
\end{split}
\end{equation}

Whenever the effective cosmological constant is non-vanishing, it is crucial that the masses satisfy $m^2\geq 2\bar{\Lambda}/(D-1)$. This is the well-known Higuchi bound, below which the helicity-0 graviton modes become ghost-like \cite{OG_Higuchi,OG_Higuchi_2,Higuchi_bound_massive_grav,General_Higuchi_bound}. It is a sufficient condition that the lightest massive mode exceeds this bound, since all the heavier modes then will do too. 

At the point where the Higuchi bound is saturated, the helicity-0 component of the corresponding mass mode becomes pure gauge and so the number of propagating degrees of freedom is reduced by 1: this is the linear `partially massless' (PM) theory \cite{Gauge_inv_vs_massless,Gauge_inv_phases_PM,PM_higher_spin,PM_at_Higuchi,Null_prop_PM,conformal_inv_PM}. The PM theory has been the subject of much interest in the context of bigravity, since many of the linear instabilities that exist when all graviton degrees of freedom propagate disappear if the one massive mode exhibits PM invariance (see e.g. \cite{PM_vs_BH}). However, there is strong evidence that the PM gauge invariance does not survive to the full non-linear level within the dRGT framework \cite{no_nonlinear_PM,dR_no_nonlinear_PM,nogo_PM,no_2deriv_action_PM}, and in the general multi-metric theory even at the linear level only the lightest mass mode can exhibit PM invariance anyway, so any instabilities present still exist for the heavier modes (although it should be stated that a nonlinear theory of interacting PM spin-2 fields does exist within the realm of conformal gravity \cite{looking_for_PM,theory_nonlin_PM,interactions_PM}).

As a sanity check, one may compare the results of this section against known results in bigravity, where the mass matrix is simple enough to diagonalise explicitly. In bigravity, one has $N=2$ metrics, and usually denotes $a_0=1, a_1=c$. Only $\Sigma_0^{(+)}$ and $\Sigma_1^{(-)}$ are present in the mass matrix, and by Eq. \eqref{Sig_pm} these are related as $\Sigma_1^{(-)}=c^{2-D}\Sigma_0^{(+)}$. Therefore, one has explicitly that:
\begin{equation}
    \mathcal{M}^2 =
    \begin{bmatrix}
        M_0^{-(D-2)}c\Sigma_0^{(+)} & -(M_0M_1)^{-\frac{D-2}{2}}\Sigma_0^{(+)}
        \\
        -(M_0M_1)^{-\frac{D-2}{2}}c^{4-D}\Sigma_0^{(+)} & M_1^{-(D-2)}c^{3-D}\Sigma_0^{(+)}
    \end{bmatrix}
\end{equation}
which one can check has eigenvalues:
\begin{align}
    m_0^2 &= 0 \; ,
    \\
    m_1^2 &= \frac{c\Sigma_0^{(+)}}{M_0^{D-2}}\left[\frac{1+(\gamma c)^{D-2}}{(\gamma c)^{D-2}}\right] \; ,
\end{align}
with the corresponding mass modes being:
\begin{align}
    H^{(0)}_{\mu\nu} &= \frac{1}{\sqrt{1+c^2\gamma^{D-2}}}\left(h^{(0)}_{\mu\nu} + c \gamma^{\frac{D-2}{2}}h^{(1)}_{\mu\nu}\right) \; ,
    \\
    H^{(1)}_{\mu\nu} &= \frac{1}{\sqrt{1+c^{2(D-3)}\gamma^{D-2}}}\left(h^{(1)}_{\mu\nu} -c^{D-3}\gamma^{\frac{D-2}{2}}h^{(0)}_{\mu\nu}\right) \; ,
\end{align}
defining $\gamma=M_{1}/M_{0}$ as the ratio of the gravitational couplings. The equations of motion for the mass modes, as well as the corresponding Fierz-Pauli mass $m_1$, agree precisely with the bigravity results \cite{consistent_spin2,Bigravity_gen_dim}, as they should (though in \cite{Bigravity_gen_dim} the perturbations were defined differently to in our Eq. \eqref{expanded metric}, as $a_i^2\bar{g}_{\mu\nu}+h^{(i)}_{\mu\nu}$, without the extra factors of $a_i$ and $M_i$, so the precise structure of their $H^{(i)}_{\mu\nu}$ in terms of the $h^{(i)}_{\mu\nu}$ is slightly different to here -- see appendix \ref{App:linearisation} for more details).

As before, there are many special cases that are interesting to consider in their own right, within this general framework. A couple of illustrative examples in $D=4$ are again useful to demonstrate, before we continue to investigate the stability of the black hole solutions. First, we return to the clockwork scenario.

\subsubsection{Clockwork gravity in 4d}

We defined a clockwork theory back in section \ref{Sec:BS} as a member of a particular class of multi-metric theories that admit a vacuum structure with an exponential hierarchy i.e. $a_i=a_0/q^{i}$ for $q\gtrsim1$, the idea being that we would like to generate a suppressed coupling of matter to the massless mode without introducing such hierarchies in the parameters of the theory. Here, we shall see how this works explicitly.

The simplest $D=4$ model one may construct that admits the desired vacuum structure is the single scale model of \cite{ClockworkCosmo}, which takes all $M_i=M$ and has potential coefficients given by $\beta_m^{(i,i+1)}=\alpha_i\beta_m$, where $\alpha_{-1}=\alpha_{N-1}=0$, and\footnote{Actually, there was a sign error in the mass matrix in our previous work \cite{ClockworkCosmo}, and we have also been more careful to match the normalisation of the corrected mass matrix, given by Eqs. \eqref{Mii} and \eqref{Mi1}, to that of the kinetic term. This means that the single scale model we refer to now actually had gravitons with negative mass-squared (the deconstructed RS model also in \cite{ClockworkCosmo} was fine), although the work there was only focussed on the \emph{background} cosmology so this did not come into play in the corresponding calculations. Regardless, we shall fix the error here and flip the sign of the potential coefficients so that the masses are positive.} 
\begin{itemize}
    \itemsep0em
    \item $\alpha_i = M^4 \;\;\; \forall\,i\neq-1,N-1$
    \item $\beta_0 = -6q^{-1}$
    \item $\beta_1 = 3$
    \item $\beta_2 = -q$
    \item $\beta_3 = \beta_4 = 0 \; .$
\end{itemize}
With these choices, the mass matrix has components:
\begin{align}
    \mathcal{M}_{ii}^2 &= \frac{a_0^2 q^{-2i}}{M^2}\left(\alpha_i q^{-1}+\alpha_{i-1}q^3\right) \; ,
    \\
    \mathcal{M}_{i+1,i}^2=\mathcal{M}_{i,i+1}^2 &= -\frac{a_0^2q^{-2i}}{M^2} \alpha_i \; ,
\end{align}
in agreement with \cite{ClockworkCosmo} (up to the minus sign and numerical factor we have corrected here). Numerically, the lightest mass mode is found to roughly have mass $m_1\sim qM$, with the heavier modes distributed exponentially above this. The massless eigenvector is given by $O^{j0}=\mathcal{N}/q^j$, where the normalisation is
\begin{equation}
    \mathcal{N} = \frac{1}{\sqrt{\sum_{i=0}^{N-1} q^{-2i}}} = \sqrt{\frac{1-q^{-2}}{1-q^{-2N}}} \; .
\end{equation}

If matter couples only to the metric $g^{(N-1)}_{\mu\nu}$ at the end of the chain of interactions i.e. only $\bar{T}^{(N-1)}_{\mu\nu}$ is non-vanishing (to engineer the greatest possible suppression of scales), and one chooses to fix the overall normalisation of the conformal factors such that $a_{N-1}=1$, then it follows that the massless mode has the following dynamics:
\begin{equation}
    \bar{\mathcal{E}}^{\alpha\beta}_{\;\mu\nu}H^{(0)}_{\alpha\beta} = \frac{1}{M_{\text{Pl}}} \bar{T}^{(N-1)}_{\mu\nu} \; ,
\end{equation}
where the effective Planck scale is:
\begin{equation}
    M_{\text{Pl}} = \sqrt{\frac{1-q^{-2N}}{1-q^{-2}}}q^{N-1} M \; .
\end{equation}
This can be much larger than $M$ if the number of fields in the chain is big enough -- indeed, generating this scale hierarchy is precisely the purpose of the clockwork mechanism.

\subsubsection{Deconstructed flat extra dimension}

Another example, which is interesting to consider because it is simple enough to analyse analytically, is the multi-metric model arising from the deconstruction of a flat extra dimension. As we saw in section \ref{Sec:BS}, the 4D multi-metric (proportional) solutions correspond to dimensionally deconstructed solutions of some 5D gravitational theory (generically a scalar-tensor braneworld), with the geometry of the extra dimension encoded in the vacuum structure of the conformal factors. We saw this explicitly for the black string solution, but the result is true more generally i.e. for any $\bar{g}_{\mu\nu}$ satisfying Einstein's equations. For example, the clockwork vacuum structure Eq. \eqref{VacStructure} corresponds to an extra dimension that is warped (as there is exponential damping through the chain of hypersurfaces), though one could also consider the simpler situation where all $a_i=1$, corresponding to an extra dimension that is flat. 

In particular, one may choose their model parameters in such a way that the 4D theory, in its continuum limit, becomes pure 5D GR without cosmological constant. Explicitly, the model in question has again equivalent 4D gravitational couplings $M_i=M_{(4)}$, as well as potential coefficients given again by $\beta_m^{(i,i+1)}=\alpha_i\beta_m$ with $\alpha_{-1}=\alpha_{N-1}=0$, except this time \cite{Deconstructing,ClockworkCosmo}:
\begin{itemize}
    \itemsep0em
    \item  $\alpha_i = M^3_{(5)}/\delta y \;\;\; \forall \,i\neq0,N-1$
    \item $\beta_0 = -6$
    \item $\beta_1 = 3$
    \item $\beta_2 = -1$
    \item $\beta_3 = \beta_4 = 0 \; ,$
\end{itemize}
where $\delta y$ is the spacing of the hypersurfaces upon which the multi-gravity metrics live in the extra dimension, and $M_{(5)}$ is the 5D gravitational coupling, related to the 4D coupling as $M^2_{(4)}=M^3_{(5)}\delta y$. With these parameters, one recovers 5D GR from the 4D multi-metric theory upon taking the limit where $\delta y\rightarrow 0$ and $N\rightarrow\infty$ while keeping the product $N\delta y = L$ fixed (corresponding to the size of the extra dimension). One may check that substituting these parameters into the $D=4$ multi-metric vacuum equations \eqref{vac_eqs} gives a solution with $\bar{\Lambda}=0$ and $a_i=1\;\forall\,i$.

The form of the mass matrix here reflects the simplicity of the vacuum, reading:
\begin{equation}
    \mathcal{M}^2 = \frac{1}{\delta y^2}
    \begin{bmatrix}
       1 & -1 & 0 & 0 & \dots & 0 & 0
       \\
       -1 & 2 & -1 & 0 & \dots & 0 & 0
       \\
       0 & -1 & 2 & -1 & \dots & 0 & 0
       \\
       0 & 0 & -1 & 2 & \dots & 0 & 0
       \\
       \vdots & \vdots & \vdots & \vdots & \ddots & \vdots & \vdots
       \\
       0 & 0 & 0 & 0 & \dots & 2 & -1
       \\
       0 & 0 & 0 & 0 & \dots & -1 & 1
    \end{bmatrix}_{(N\times N)} \; ,
\end{equation}
which one can diagonalise using the techniques developed in \cite{Losonczi}. The result is that the mass eigenvalues are given by:
\begin{equation}
    m_n^2 = \frac{4}{\delta y^2}\sin^2\left(\frac{n\pi}{2N}\right) \; ,
\end{equation}
the massless mode is simply:
\begin{equation}
    H^{(0)}_{\mu\nu} = \frac{1}{\sqrt{N}}\sum_{n=0}^{N-1} h^{(n)}_{\mu\nu} \; ,
\end{equation}
while the massive modes are:
\begin{equation}
    H^{(m)}_{\mu\nu} = \mathcal{N}_m \sum_{n=0}^{N-1}\left[\sin\left(\frac{(n+1)m\pi}{2N}\right) - \sin\left(\frac{nm\pi}{2N}\right)\right] h^{(n)}_{\mu\nu} \; ,
\end{equation}
with normalisations $\mathcal{N}_m$ whose explicit expressions are irrelevant but can easily be determined.

The mass eigenvalues and structure of the massless mode take the expected functional forms \cite{deconstructing_dims,SC_and_graph_structure}, with the masses of the lightest modes scaling as $m_n\sim n/L$ and the heaviest as $m_{N-1}\sim N/L$. Similarly, if one again couples matter to one of the metrics, then one recovers the expected relations between the effective Planck scale of the massless mode and the various gravitational couplings i.e. $M_{\text{Pl}}^2=NM_{(4)}^2=NM_{(5)}^3\delta y=M^3_{(5)}L$.

\section{Linear stability of Proportional Solutions}\label{Sec:stability}

Now we have all we need to discuss the linear stability of the proportional black hole solutions. First, we note that the linearised version of the Bianchi constraint in vacuum implies that $\bar{\nabla}^\nu H^{(i)}_{\mu\nu}=\bar{\nabla}_\mu H^{(i)}$, and so $\bar{\nabla}^\nu\bar{\nabla}^\mu H^{(i)}_{\mu\nu}=\bar{\Box}H^{(i)}$. Using the latter in the trace of Eqs. \eqref{linearised eqs}, one is forced into de Donder gauge for \emph{all} the mass modes simultaneously:
\begin{equation}\label{TT gauge}
    \bar{\nabla}^\nu H^{(i)}_{\mu\nu} = H^{(i)} = 0 \;\;\; \forall i \; .
\end{equation}
With this restriction, the linearised equations for the mass modes become:
\begin{equation}\label{TT equations}
    \bar{\Box}H^{(i)}_{\mu\nu} + 2\bar{R}^{\alpha\;\beta}_{\;\mu\;\nu}H^{(i)}_{\alpha\beta} = m_i^2 H^{(i)}_{\mu\nu} \; ,
\end{equation}
whose behaviour can then be analysed depending on one's choice for the background metric $\bar{g}_{\mu\nu}$.

\subsection{Multi-Schwarzschild}

Eqs. \eqref{TT equations}, with the common background $\bar{g}_{\mu\nu}$ taken as the $D=4$ Schwarzschild metric, are precisely those equations studied in the context of the GL instability \cite{GL_instability,Charged_GL_instability,AdS_GL_instability,GL_instability_bigravity,GL_unified,Kerr_instability_bigravity}. In the original work, linear perturbations to the 5D black string were considered, and split into scalar, vector and tensor contributions \`a la Kaluza-Klein. Fourier decomposing around the extra dimension, it was shown that the 4D tensor perturbations (i.e. the Fourier coefficients of the tensor contribution) satisfy precisely Eqs. \eqref{TT equations}, just with the corresponding Fourier momentum in place of the mass. Spherically symmetric $s$-wave solutions, regular at the future event horizon, were found of the form:
\begin{equation}
    h^{(4)}_{\mu\nu} = e^{\Omega t}
    \begin{bmatrix}
        h_{tt}(r) & h_{tr}(r) & 0 & 0
        \\
        h_{rt}(r) & h_{rr}(r) & 0 & 0
        \\
        0 & 0 & K(r) & 0
        \\
        0 & 0 & 0 & K(r)\sin^2{\theta}
    \end{bmatrix} \; ,
\end{equation}
and were shown to be unstable (i.e. have $\Omega>0$) within the range:
\begin{equation}\label{mass stability bound}
    0 < m < \mathcal{O}\left(\frac{1}{r_s}\right) \; .
\end{equation}

In \cite{GL_instability_bigravity}, it was argued in bigravity that since the one massive mode satisfies the same equations as the 4D tensor perturbations in the context of the black string, the bi-Schwarzschild solutions in bigravity too are unstable if the mass satisfies the inequality \eqref{mass stability bound}. This idea was made more concrete in \cite{Kerr_instability_bigravity}, who studied the dynamics of the unstable mode in detail for bi-Schwarzschild-dS, and determined that the instability turns on when $mr_s\lesssim 0.86$ (which, interestingly, is precisely the threshold at which the hairy solutions discussed briefly in section \ref{Sec:Hairy} come into existence \cite{Hairy_BHs_flat}, signalling that the hairy black holes may actually be the end point of the instability -- for a nice review of this situation see \cite{BHs_review}).

Since the mass modes in the multi-metric theory all independently obey Eqs. \eqref{TT equations} for their respective mass eigenvalues, it is clear how the arguments of \cite{GL_instability_bigravity,Kerr_instability_bigravity} generalise naturally: the multi-metric theory possesses a single massless mode and a tower of massive modes, whose masses are given in terms of the parameters of the theory through the mass matrix $\mathcal{M}^2$; \emph{all} of the mass modes are subject to the same stability condition, where the proportional Schwarzschild solution becomes unstable if, for any of the $m_i$, $m_i r_s\lesssim0.86$. Consequently, the solution is stable if and only if the lightest massive mode $H^{(1)}_{\mu\nu}$ sits above this inequality. This is for the simple reason that if the lightest mass mode evades the instability then necessarily so too do all of the others. Conversely, unstable solutions may have multiple unstable mass modes.

In principle, one could construct a multi-metric theory with whatever masses one wishes, depending on one's choice for the potential coefficients. In practice, however, such choices are tightly constrained, particularly by Solar System tests of gravity, as the additional degrees of freedom associated with the mass modes can induce marked deviations from GR even in the weak-field regime \cite{GR_tests}. Typically, the masses have been taken of order the Hubble parameter today ($m\sim H_0\sim 10^{-33}\text{eV}$) with the goal of addressing the question of dark energy; the Vainshtein mechanism then ensures that Solar System tests are satisfied, as GR is restored due to nonlinear self-interactions of the helicity-0 graviton modes \cite{Vainshtein,Vainshtein_decoupling_limit,Vainshtein_recovery,Restoring_GR_bigrav,Vainshtein_review}. However, this choice is not a necessary one stemming from any sort of theoretical or observational requirement; in particular, a multi-metric theory whose masses are all very heavy will also restore GR at the linear level, without the need for any Vainshtein screening, since the Compton wavelengths $m^{-1}$ of the mass modes are small enough that their effects would be invisible to current experimental precision. More precisely, modifications to weak-field GR solutions (at least in bigravity) are suppressed by a factor $e^{-mr}$ \cite{Hairy_BHs_flat}, implying that for large $m$ the large scale behaviour in the Solar System should be indistinguishable from GR\footnote{This point, together with the fact that the heavy spin-2 still gravitates in the same way as ordinary matter, was used (in bigravity) in \cite{bigravity_DM,heavy_spin2_DM} to argue that the heavy mode could also constitute an interesting dark matter candidate.}. This is possible due to the surviving massless mode in the multi-metric theory -- in dRGT massive gravity, where the only propagating graviton is massive, large masses are ruled out, as predictions always differ from GR due to the vDVZ discontinuity \cite{vD,Zakharov}.

The black hole instability affects the two different mass regimes in different ways. When the graviton masses are ultra-light, by virtue of Eq. \eqref{mass stability bound} all astrophysically relevant black holes are unstable: if $m\sim H_0$ then any Schwarzschild black hole weighing less than $10^{22}M_\odot$ suffers from the instability \cite{BHs_review}. However, for such light graviton masses, $\Omega$ scales linearly with $m$ \cite{blackfold_dynamics,brane_viscosity,AdS_RF_correspondence_GL}, so the characteristic timescale of the instability $\Omega^{-1}$ is of order the Hubble time: while the instability may always be present, it is not physically relevant over any observable timescale. On the other hand, if the graviton masses are very heavy, then while the instability is far more efficient, it affects only the very lightest black holes. For example, if the lightest graviton lies at the TeV scale (which is optimal for a dark matter candidate \cite{bigravity_DM,heavy_spin2_DM} and is also sensible in clockwork scenarios \cite{ClockworkGrav,ClockworkCosmo}), only black holes weighing less than roughly $10^{-22}M_\odot$ are unstable. However, the instability timescale is now much shorter, and potentially \emph{is} physically relevant -- some primordial black holes, for example, may exist in the unstable mass range, and as initially stable black holes evaporate by Hawking radiation, they will inevitably cross over into the unstable regime at some time. It is difficult to say much more than this without knowing for definite the end state of the instability (which requires a full nonlinear analysis, though one possibility is the hairy solution of \cite{Hairy_BHs_flat}). However, it may be possible that such effects possess observational signatures, or even signal at pathologies within the multi-metric theory itself.

\subsection{Multi-Kerr}

If one now allows the black holes to rotate, and so instead takes the common background $\bar{g}_{\mu\nu}$ as the Kerr metric, then the GL monopolar instability is still present \cite{Kerr_instability_bigravity}, but the value of $m_ir_s$ for which modes become stable increases relative to the Schwarzschild case ($m_ir_s\sim0.86$) with increasing black hole spin \cite{backreaction} (though it still always remains order 1). There are also additional instabilities associated with the azimuthal modes that are not present in the Schwarzschild case, as a consequence of the superradiant instability of rotating black holes against massive bosonic excitations \cite{floating_orbits,KG_rotating_BH,superradiance2020}. This effect occurs when the frequency of the perturbation satisfies:
\begin{equation}\label{superrad condition}
    0<\omega<m_{A}\Omega_{\text{BH}} \; ,
\end{equation}
where $m_{A}$ is the mode's azimuthal quantum number and $\Omega_{\text{BH}}$ is the angular velocity of the black hole horizon, and is characterised by the bosons forming a condensate around the black hole, which then spins down and deposits its rotational energy into the condensate until the above bound saturates at $\omega=m_A\Omega_{\text{BH}}$. The condensate then dissipates via (almost monchromatic) quadrupolar gravitational wave emission \cite{BH_detectors}. 

The superradiant instability described above turns out to be most effective when the Compton wavelength of the perturbation in question is comparable with the horizon size of the black hole \cite{superradiance_ULS2,superradiance_massive_spin2}. Therefore, like the GL monopolar instability ($m_A=0$), it is relevant for black holes that have $m_ir_s\sim\mathcal{O}(1)$. Unlike the GL instability, however, this is not a hard bound at which the superradiant instability switches on, rather a statement on when its rate is fastest -- all black holes with rotation velocities satisfying Eq. \eqref{superrad condition} suffer, but the instability rate will only be non-negligible for a certain range of black hole masses (given a value for $m_i$). Consequently, in the multi-metric theory, a wider range of black hole masses will be affected than would be with a single massive graviton, with successively heavier mass modes having potentially relevant superradiant instabilities for successively lighter black holes ($m_ir_s$ being order 1 corresponds to $m_i\sim10^{-11}(M_\odot/M_{\text{BH}})$ eV, in physical units). 

The rates of the superradiant instabilities for massive spin-2 fields have been studied semi-analytically for small black hole spins in \cite{Kerr_instability_bigravity}, analytically in the regime where $m_ir_s\ll1$ in \cite{superradiance_ULS2}, and fully numerically for $m_ir_s$ up to 0.8 and spins up to $j=0.99$ in \cite{superradiance_massive_spin2}, where it was found that the dipole ($m_A=1$) mode is the fastest growing -- for certain regions of parameter space it is so fast that it can even affect black hole ringdown. The authors as a result argued that a measurement of non-zero rotation in supermassive black holes would rule out large swathes of ultra-light graviton masses. However, it was demonstrated in \cite{backreaction} that in most of the parameter space (save for the very fastest spins and largest $m_ir_s$) the growth of the superradiant instabilities is always subdominant to that of the GL monopolar mode, so any such constraints must take into account the backreaction of the GL mode on the solution. Again, to do this, a nonlinear analysis with knowledge of the end state of this instability for Kerr black holes will be necessary. This of course requires one to have a well-posed dynamical formulation of the multi-metric theory that is suitable for such simulations, upon which the development is ongoing (see \cite{Dynamical_dRGT} for the case of dRGT massive gravity with flat reference metric). That said, as an initial toy model for the nonlinear evolution of the instability, the authors of \cite{backreaction} considered the linearised system as arising from Einstein-Weyl theory \cite{Stelle1,Stelle2}, which has a well-posed dynamical formulation \cite{IVP_stelle,Cauchy_Stelle,Nonlin_stelle}, but also contains a ghost. They nevertheless found signatures of the linear theory in their nonlinear analysis, and postulated that this may be a general feature of all such nonlinear analyses i.e. even in the ghost free nonlinear theories, including the multi-metric one.

\section{Conclusion}\label{Sec:conclusion}

In this work, we have sought to extend and generalise numerous results regarding 4-dimensional black holes in the theories of dRGT massive gravity and bigravity to the general ghost free multi-metric theory in arbitrary dimension. To that end, we have explicitly constructed various example black hole spacetimes that solve the multi-metric equations of motion, including analogues in the proportional branch of all the higher dimensional Myers-Perry black holes of GR, as well as multiple additional solutions in which not all metrics are simultaneously diagonalisable (including a class of solutions -- the partially proportional branch -- which is not present in dRGT/bigravity). The additional solutions we constructed describe, respectively: asymptotically (A)dS rotating black holes in arbitrary dimension, asymptotically (A)dS charged black holes in arbitrary dimension, as well as asymptotically (A)dS charged \emph{and} rotating black holes in $D=4$ (also in $D=3$, although we did not write it down explicitly). We suspect that the hairy black hole solutions in bigravity also carry across to the multi-metric theory in a natural way, despite not performing the explicit calculation here. Furthermore, we related these multi-metric black hole solutions to the well-known black string solutions of higher dimensional GR, with the structure of the conformal factors that defines the multi-metric vacuum encoding information about the geometry of the extra dimension (we used the example of clockwork gravity to represent a warped extra dimension in the dimensional deconstruction limit, for example).

We later studied the linear stability of these multi-metric black holes. After linearising the general theory to determine the dynamics of the spin-2 mass modes, we showed that the GL and superradiant instabilities that plague 4-dimensional proportional black hole solutions in dRGT massive gravity and bigravity carry over naturally to the multi-metric theory (as they should, given the relation to black strings). More precisely: in dRGT/bigravity, in terms of the one graviton mass, it is known that a bound exists at which the GL instability turns on, as well as a regime for which the superradiant instability is most efficient, both occuring when $m r_s\sim\mathcal{O}(1)$ (differing slightly between the Schwarzschild and Kerr cases). In the multi-metric theory, this relation holds for \emph{every} graviton in the spectrum, which translates to a stability bound on only the lightest massive state for the GL mode, as well as a wider array of black hole masses that are potentially affected significantly by the superradiant modes, relative to the situation in dRGT/bigravity. 

Since observations favour either very light or very heavy graviton masses (in order for the multi-metric theory to agree with GR), the consequences of these instabilities can be vastly different, and can affect vastly varying sizes of black hole depending on the particular theory one chooses to work with (i.e. depending on a particular choice of interaction coefficients and number of metrics). In order to pin these consequences down, more work is required to understand how the instabilities saturate. This will inevitably rely on us having well-posed dynamical simulations within the framework of ghost free multi-gravity, upon which the development is ongoing. Nevertheless, we hope that as a result of this work we are now a small step closer to understanding such interesting questions.

\section*{Acknowledgements}

K.W. is supported by a UK Science and Technology Facilities Council studentship. P.M.S. and A.A. are supported by a STFC Consolidated Grant [Grant No. ST/T000732/1]. Some calculations involving various vielbein contractions, and checks thereof, were aided by use of the \emph{xAct} Mathematica package suite \cite{xAct} (specifically \emph{xCoba}). For the purpose of open access, the authors have applied a Creative Commons Attribution (CC BY) licence to any Author Accepted Manuscript version arising.

\section*{Data Access Statement}
No new data were created or analysed in this study.

\appendix
\section{Derivation of the field equations}\label{App:eqs}

To derive the vielbein form of the field equations as given in section \ref{Sec:Review}, we determine the variation of the action, Eq. \eqref{MultigravAction}, using the differential form framework prescribed in \cite{Varying_actions}. For an equivalent derivation in the metric formalism, we refer the reader to e.g. \cite{Bigravity_gen_dim}.

The variation of the kinetic term with respect to the $i$-th tetrad gives:
\begin{align}
    \delta S_K &= \frac{M_{i}^{D-2}}{2} \int_{\mathcal{M}_D} \delta e^{(i)a}\wedge R^{(i)}_{bc}\wedge \hodge^{(i)} e^{(i)bc}_{\;\;\;\;\;\;\;a}
    \\
    &= -M_{i}^{D-2}\int_{\mathcal{M}_D} \delta e^{(i)a}\wedge \hodge^{(i)} G^{(i)}_a \; ,
\end{align}
where $\hodge G^a = -\frac12 R_{bc}\wedge\hodge e^{abc}$ is the Hodge dual of the Einstein (vector-valued) 1-form whose components are those of the usual Einstein tensor i.e. $G^a=G^a_{\;b}e^b$. To see that this is indeed just the usual Einstein tensor, consider the component expression for the dual (a $(D-1)$-form):
\begin{equation}
    \hodge G^a = -\frac{1}{4(D-3)!} R_{bcmn}\varepsilon^{abc}_{\;\;\;\;d\hdots} e^{mnd\hdots} \; ,
\end{equation}
from which we can use the fact that $e_k\wedge\hodge G^a = G^a_{\;k}\hodge1$ to extract the components. First, we have:
\begin{align*}
    e_k\wedge\hodge G^a &= -\frac{1}{4(D-3)!}R_{bcmn}\varepsilon^{abc}_{\;\;\;\;d\hdots}e_k^{\;mnd\hdots}
    \\
    &= +\frac{1}{4(D-3)!}R^{mn}_{\;\;\;\;bc}\varepsilon^{abcd\hdots}\varepsilon_{kmnd\hdots}\hodge1
    \\
    &= -\frac14 R^{mn}_{\;\;\;\;bc}\delta^{abc}_{kmn}\hodge1 \; ,
\end{align*}
using the symmetries of the Riemann tensor and the fact that the epsilon tensors sum to
\begin{equation}
\varepsilon_{\mu_1\mu_2\mu_3\hdots}\varepsilon^{\nu_1\nu_2\nu_3\hdots} = (-1)^s\delta^{\nu_1\nu_2\nu_3\hdots}_{\mu_1\mu_2\mu_3\hdots} \; ,
\end{equation}
for metric of signature $(s,D-s)$, where $\delta^{\nu_1\nu_2\nu_3\hdots}_{\mu_1\mu_2\mu_3\hdots}$ is the generalised Kronecker delta defined by
\begin{equation}
    \delta^{\nu_1\hdots\nu_p}_{\mu_1\hdots\mu_p} = p!\delta^{\nu_1}_{[\mu_1}\hdots\delta^{\nu_p}_{\mu_p]}  \; .
\end{equation}
We see that the Einstein tensor has (tetrad basis) components given by:
\begin{equation}\label{Einstein tensor}
    G^a_{\;k} = -\frac14 R^{mn}_{\;\;\;\;bc} \delta^{abc}_{kmn} \; ,
\end{equation}
or in more familiar form:
\begin{equation}
\begin{split}
    G^a_{\;k} &= -\frac14 R^{mn}_{\;\;\;\;bc} 3!\delta^a_{[k}\delta^b_m\delta^c_{n]}
    \\
    &= -\frac14 \left(-4R^a_{\;k} + 2\delta^a_k R\right)
    \\
    &= R^a_{\;k} - \frac12 \delta^a_k R \; .
\end{split}
\end{equation}

Now, for the potential term, the variation is:
\begin{align}
    \delta S_V
    &= -\int_{\mathcal{M}_D} \delta e^{(i)a}\wedge \hodge^{(i)} W^{(i)}_a \; ,
\end{align}
in which we define the $W$-tensor 1-form in an analogous way to our Einstein 1-forms above. Explicitly,
\begin{align}\label{W form}
    \hodge^{(i)}&W^{(i)}_a = \varepsilon_{ab_1\hdots b_{D-1}} \\&\times \sum_{j_1\hdots j_{D-1}} \mathcal{P}(i) T_{ij_1\hdots j_{D-1}} e^{(j_1)b_1} \wedge \hdots \wedge e^{(j_{D-1})b_{D-1}} \;\nonumber .
\end{align}
To extract the components, we use the same trick as before -- namely, that $\dd x^\mu \wedge \hodge^{(i)} W^{(i)}_a=W^{(i)\mu}_{\;\;\;\;\;\;a}\hodge^{(i)}1$, only this time wedging with the coordinate basis 1-form $\dd x^\mu$, since the tetrads in Eq. \eqref{W form} do not necessarily belong to the same geometry (in order to make the calculation tractable, we need to use the vielbeins to express everything in a common basis in which we can then identify the volume form $\hodge^{(i)}1$; we make the natural choice of the coordinate basis). The result is that:
\begin{widetext}
    \begin{align*}
        W^{(i)\mu}_{\;\;\;\;\;\;a}\hodge^{(i)}1 &= \varepsilon_{ab_1\hdots b_{D-1}} \sum_{j_1\hdots j_{D-1}} \mathcal{P}(i) T_{ij_1\hdots j_{D-1}}\vbein{j_1}{\lambda_1}{b_1}\hdots\vbein{j_{D-1}}{\lambda_{D-1}}{b_{D-1}}\dd x^{\mu\lambda_1\hdots\lambda_{D-1}}
        \\
        &= -\varepsilon^{\mu\lambda_1\hdots\lambda_{D-1}}\varepsilon_{ab_1\hdots b_{D-1}}\sum_{j_1\hdots j_{D-1}} \mathcal{P}(i) T_{ij_1\hdots j_{D-1}}\vbein{j_1}{\lambda_1}{b_1}\hdots\vbein{j_{D-1}}{\lambda_{D-1}}{b_{D-1}}\hodge^{(i)}1
        \\
        &= D!\ivbein{i}{\mu}{[a}\ivbein{i}{\lambda_1}{b_1}\hdots\ivbein{i}{\lambda_{D-1}}{b_{D-1}]} \sum_{j_1\hdots j_{D-1}} \mathcal{P}(i) T_{ij_1\hdots j_{D-1}} \vbein{j_1}{\lambda_1}{b_1}\hdots\vbein{j_{D-1}}{\lambda_{D-1}}{b_{D-1}}\hodge^{(i)}1 \; .
    \end{align*}
\end{widetext}
Contracting with $e^{(i)a}_{\nu}$ gives us the appropriate coordinate basis expression, Eq. \eqref{W tensor}.

Finally, the variation of the matter action defines the energy-momentum 1-form as:
\begin{equation}\label{delta SM}
    \delta S_M = \int_{\mathcal{M}_D} \delta e^{(i)a} \wedge \hodge^{(i)} T^{(i)a} \; .
\end{equation}
Putting the three variations together and taking the functional derivative with respect to $\delta e^{(i)a}$ gives us the differential form version of our Einstein equations:
\begin{equation}
    \hodge^{(i)} \left(  - M^{D-2}_{i}G^{(i)}_a - W_a^{(i)} + T_a^{(i)}\right) = 0 \; ,
\end{equation}
which in coordinate basis components becomes precisely the equations of motion \eqref{Einstein eqs}.

\section{Derivation of the linearised field equations}\label{App:linearisation}

This time, the calculation is simpler in the metric formalism, and follows closely the bigravity derivation in \cite{Bigravity_gen_dim}. First, recall that the metric form of the $W$-tensor for chain type interactions reads (C.F. Eq. \eqref{metric W}):
\begin{equation}\label{metric W_app}
\begin{split}
    \Wi &= \sum_{m=0}^D(-1)^m\beta_m^{(i,i+1)}Y_{(m)\nu}^\mu(S_{i\rightarrow i+1})
    \\
    &+\sum_{m=0}^D (-1)^m\beta_{D-m}^{(i-1,i)}Y_{(m)\nu}^\mu (S_{i\rightarrow i-1}) \; .
\end{split}
\end{equation}
 
To see that this is entirely equivalent to the correpsonding vielbein formalism result, Eq. \eqref{W_comps_prop}, notice that the building block matrices with the proportional ansatz ($\gi=a_i^2\bar{g}_{\mu\nu}$) take the simple form $S_{i\rightarrow i\pm1}=(a_{i\pm1}/a_i)\mathbbm{1}$, which means that the eigenvalues of $S_{i\rightarrow i\pm1}$ are simply $D$ copies of $(a_{i\pm1}/a_i)$. Therefore, the elementary symmetric polynomials are:
\begin{equation}
    e_k(S_{i\rightarrow i\pm1})=\left(\frac{a_{i\pm1}}{a_i}\right)^k\binom{D}{k} \; .
\end{equation}
Substituting into Eq. \eqref{Y_def}, and using the binomial coefficient identity $\sum_{k=0}^m(-1)^k\binom{D}{k}=(-1)^m\binom{D-1}{m}$, one finds that:
\begin{equation}
    Y_{(m)}(S_{i\rightarrow i\pm1}) = (-1)^ma_{i\pm1}^m a_i^{-m}\binom{D-1}{m}  \mathbbm{1}  \; ,
\end{equation}
which recovers Eq. \eqref{W_comps_prop} upon substitution back into Eq. \eqref{metric W_app}.

To linearise the system, we perturb around our proportional background. The metric and its inverse are expanded as:
\begin{align}
    \gi &= a_i^2\bar{g}_{\mu\nu} + \delta g^{(i)}_{\mu\nu}\label{dg1}
    \\
    g^{(i)\mu\nu} &= a_i^{-2}\bar{g}^{\mu\nu} - a_i^{-4}\delta g^{(i)\mu\nu}\label{dg2} \; ,
\end{align}
where we have included factors of $a_i$ so that $\delta g^{(i)}_{\mu\nu}$ behaves as if it lived in the \emph{common} background described by metric $\bar{g}_{\mu\nu}$ (if $\delta g^{(i)}_{\mu\nu}$ had its indices instead manipulated with $a_i^2\bar{g}_{\mu\nu}$, the inverse would simply be $g^{(i)\mu\nu} = a_i^{-2}\bar{g}^{\mu\nu} - \delta g^{(i)\mu\nu}$).

It is well known (see e.g. \cite{dR_review}) that the Einstein tensor linearises to the Lichnerowicz operator acting on the perturbation, that is:
\begin{equation}
    \delta G^{(i)}_{\mu\nu} = \bar{\mathcal{E}}^{\alpha\beta}_{\;\mu\nu}\delta g^{(i)}_{\alpha\beta} \; ,
\end{equation}
so we shall skip this part of the derivation here and focus on the potential. To that end, the first order variation of the $W$-tensor is:
\begin{equation}\label{delta W}
\begin{split}
        \delta\Wi = \delta g^{(i)\mu}_{\;\;\;\;\;\lambda}&\bar{W}^{(i)\lambda}_{\;\;\;\;\;\;\nu} + \sum_{m=0}^D(-1)^m\beta_m^{(i,i+1)}\delta Y_{(m)\nu}^\mu(S_{i\rightarrow i+1})
        \\
        &+ \sum_{m=0}^D(-1)^m\beta_{D-m}^{(i-1,i)}\delta Y_{(m)\nu}^\mu(S_{i\rightarrow i-1}) \; ,
\end{split}
\end{equation}
where, around the proportional background, we know that $\bar{W}^{(i)\lambda}_{\;\;\;\;\;\;\nu}=\delta^\lambda_\nu M_i^{D-2}\bar{\Lambda}/a_i^2$ (see section \ref{Sec:prop sols}). The variation of the $Y$'s is given by \cite{linear_spin2_gen_BG}:
\begin{align}
    \delta Y_{(m)}(S) = \sum_{k=1}^m (-1)^k &\left[ S^{m-k}\delta e_k(S) \right.
    \\
    &\left. -e_{k-1}(S)\sum_{n=0}^{m-k} S^n \delta S S^{m-k-n} \right] \; ,\nonumber
\end{align}
where, by virtue of Eq. \eqref{sym pols}, we have:
\begin{equation}\label{delta ek}
    \delta e_k(S) = -\sum_{n=1}^k (-1)^n\Tr(S^{n-1}\delta S) e_{k-n}(S) \; .
\end{equation}
Since it is built from the metric variations, $\delta S_{\mu\nu}$ also has its indices manipulated with the common background metric $\bar{g}_{\mu\nu}$.

Eqs. \eqref{delta W}-\eqref{delta ek} hold in general, but around the proportional background things simplify greatly: as we saw, the building block matrices take the simple form $S_{i\rightarrow i\pm1}=(a_{i\pm1}/a_i)\mathbbm{1}$. After substituting this in above (and using some properties of the binomial coefficients) one can show that the $Y$ variations become \cite{Bigravity_gen_dim}:
\begin{align}
    \delta Y_{(m)}(S_{i\rightarrow i\pm1}) = &(-1)^m a_{i\pm1}^{m-1}a_i^{1-m}\binom{D-2}{m-1}\\&\times\left[\Tr(\delta S_{i\rightarrow i\pm1})\mathbbm{1}-\delta S_{i\rightarrow i\pm1}\right]\nonumber \; ,
\end{align}
and so the variation of the $W$-tensor is:
\begin{equation}
    \begin{split}
        \delta\Wi = \frac{M_i^{D-2}\bar{\Lambda}}{a_i^2}&\delta g^{(i)\mu}_{\;\;\;\;\;\lambda} + \Sigma_i^{(+)}\left[\delta S_{i\rightarrow i+1}\delta^\mu_\nu - (\delta S_{i\rightarrow i+1})^\mu_{\;\nu}\right]
        \\
        &+ \Sigma_i^{(-)}\left[\delta S_{i\rightarrow i-1}\delta^\mu_\nu - (\delta S_{i\rightarrow i-1})^\mu_{\;\nu}\right] \; ,
    \end{split}
\end{equation}
where $\Sigma_i^{(\pm)}$ are precisely as in Eqs. \eqref{sig_p} and \eqref{sig_m}.\pagebreak

All that remains, in order to determine the linearised equations of motion, is to calculate the precise form of $\delta S_{i\rightarrow i\pm1}$. By starting with $(S_{i\rightarrow i\pm1}^2)^{\mu}_{\;\nu}=g^{(i)\mu\lambda}g^{(i\pm1)}_{\lambda\nu}$ and substituting in Eqs. \eqref{dg1} and \eqref{dg2} for the perturbed metrics, one can show that the desired variation is given by:\pagebreak
\begin{equation}
    (\delta S_{i\rightarrow i\pm1})_{\mu\nu} = \frac{1}{2a_i^2}\frac{a_i}{a_{i\pm1}}\left[\delta g^{(i\pm1)}_{\mu\nu}-\left(\frac{a_{i\pm1}}{a_i}\right)^2\delta\gi\right] \;.
\end{equation}

With this, the linearised (vacuum) equations take the following form:
\begin{widetext}
\begin{equation}\label{linearised eqs dg}
    \begin{split}
        \bar{\mathcal{E}}^{\alpha\beta}_{\;\mu\nu}\delta g^{(i)}_{\alpha\beta} + \bar{\Lambda}\delta g_{\mu\nu}^{(i)} + \frac{a_i^2}{M_i^{D-2}}&\Bigg\{\Sigma_i^{(+)}\left[\frac{a_{i+1}}{a_i}\left(\delta g^{(i)}_{\mu\nu}-\bar{g}_{\mu\nu}\delta g^{(i)}\right) - \frac{a_i}{a_{i+1}}\left(\delta g^{(i+1)}_{\mu\nu} - \bar{g}_{\mu\nu}\delta g^{(i+1)}\right)\right]
        \\
        &+\Sigma_i^{(-)}\left[\frac{a_{i-1}}{a_i}\left(\delta g^{(i)}_{\mu\nu}-\bar{g}_{\mu\nu}\delta g^{(i)}\right) - \frac{a_i}{a_{i-1}}\left(\delta g^{(i-1)}_{\mu\nu} - \bar{g}_{\mu\nu}\delta g^{(i-1)}\right)\right]\Bigg\} = 0 \; ,
    \end{split}
\end{equation}
\end{widetext}

One can check for bigravity, where only the $i=0$ and $i=1$ terms are present, that these equations reduce to precisely the linearised equations of \cite{Bigravity_gen_dim}.

If one instead parametrises their perturbations as 
\begin{equation}
    \delta g^{(i)}_{\mu\nu}=\frac{a_i}{M_i^{\frac{D-2}{2}}} h^{(i)}_{\mu\nu} \; ,
\end{equation}
as we did in section \ref{Sec:linear}, then one recovers our linearised equations \eqref{linearised eqs}, as well as our expressions for the mass matrix components, Eqs. \eqref{Mii} and \eqref{Mi1}. The mass matrix for the $h^{(i)}_{\mu\nu}$ of course has different components to the mass matrix for $\delta g^{(i)}_{\mu\nu}$, but since their respective equations are just related by a simple field rescaling, they share the mass same eigenvalues. Therefore, the equations for the mass modes that one obtains after diagonalising are equivalent regardless of whether one initially uses $\delta g^{(i)}_{\mu\nu}$ or $h^{(i)}_{\mu\nu}$ to express their perturbations.

Finally, we note that if $\bar{g}_{\mu\nu}$ is the $D$-dimensional (A)dS metric, then one may additionally pull out the cosmological constant from the $\bar{R}^{\alpha\;\beta}_{\;\mu\;\nu}$ background curvature contribution to $\bar{\mathcal{E}}^{\alpha\beta}_{\;\mu\nu}$, to express the first two terms in Eq. \eqref{linearised eqs dg} as:
\begin{equation}
    \tilde{\mathcal{E}}^{\alpha\beta}_{\;\mu\nu}\delta g^{(i)}_{\mu\nu} -\frac{2\bar{\Lambda}}{D-2}\left(\delta g^{(i)}_{\mu\nu} - \frac12 \bar{g}_{\mu\nu}\delta g^{(i)}\right) + \hdots \; ,
\end{equation}
where $\tilde{\mathcal{E}}^{\alpha\beta}$ now contains only the covariant derivative operators, if one wishes to match up exactly with the corresponding equations as written in \cite{Bigravity_gen_dim,prop_bg_multigrav} for bigravity. For us, it is important to keep the background curvature in explicitly, since it feeds into our discussion of black hole stability, when $\bar{g}_{\mu\nu}$ is either the Schwarzschild or Kerr metric.

\section{Star type interactions}\label{App:star}

Throughout this work, we have consistently worked with chain type interactions, due to their nice interpretation as arising from dimensional deconstruction. As mentioned briefly in section \ref{Sec:Review}, however, the chain type interaction is not the only one that is devoid of the Boulware-Deser ghost: `star' type interactions, where many metrics couple to one common central metric but not to each other, are also allowed, as well as arbitrary combinations of both stars and chains provided that no interaction cycles form between the metrics. 

We would like to see how our calculations and results are altered for star type interactions. Of course, the only difference between multi-metric gravity with chain type interactions and multi-metric gravity with star type interactions lies in the structure of the interaction coefficients $\Tcoeffs$.

For the chain, only terms of the form $T_{iiii\hdots}$, $T_{i+1,iii\hdots}$, $T_{i-1,iii\hdots}$, $T_{i+1,i+1,ii\hdots}$ etc. were allowed, corresponding to the scenario where each metric interacts only with its nearest neighbours on either side. This lead us to paramterise the interaction coefficients nicely in terms of the $\beta_m^{(i,i+1)}$ of Eqs. \eqref{betas1} and \eqref{betas2}.

For the star, the situation is slightly different. If we let the index $i=0$ label the common central metric and the indices $j=(1,\hdots,N-1)$ label the outer metrics, which each couple only to $g^{(0)}_{\mu\nu}$, then the only permitted interaction coefficients are those that involve combinations of $0$ with any one distinguished $j$ e.g. $T_{0000\hdots}$, $T_{0111\hdots}$, $T_{0022\hdots}$, $T_{3333\hdots}$ and so on\footnote{One cannot allow, for example, $T_{012\hdots}$, as this would imply an interaction cycle between $g^{(0)}_{\mu\nu}$, $g^{(1)}_{\mu\nu}$ and $g^{(2)}_{\mu\nu}$, which as we said in section \ref{Sec:Review} renders the theory ghostly.}. One may parametrise the allowed coefficients in the following manner:
\begin{align}
    D!\,T_{0000\hdots 0} &= \sum_{j=1}^{N-1} \beta_0^{(0,j)} \; ,
    \\
    D!\,T_{\{j\}^m \{0\}^{D-m}} &= \beta_m^{(0,j)} \; ,
\end{align}
where $m=1,\hdots,D$, the form of the coefficients now reflecting the fact that the central $0$-th metric interacts with all $N-1$ outer metrics while the outer $j$-th metrics interact only with the central one, and that all interactions are oriented outward from the central metric. These $\beta_m^{(0,j)}$ again coincide with those of the metric formalism. With this parametrisation, one can substitute into Eq. \eqref{W tensor} to determine the form of the $W$-tensors for a given set of vielbeins. 

For the proportional metric ansatz, the result is that the $0$-th $W$-tensor has components:
\begin{equation}
    W^{(0)\mu}_{\;\;\;\;\;\;\nu} = \delta^\mu_\nu \sum_{j=1}^{N-1}\sum_{m=0}^{D} \beta_m^{(0,j)}\binom{D-1}{m} a_{j}^m a_0^{-m} \; ,
\end{equation}
while the corresponding expression for the $j$-th $W$-tensor is:
\begin{equation}
    W^{(j)\mu}_{\;\;\;\;\;\;\;\;\nu} = \delta^\mu_\nu \sum_{m=0}^{D} \beta_{D-m}^{(0,j)}\binom{D-1}{m} a_0^m a_{j}^{-m} \; .
\end{equation}
Again, the same arguments regarding the solvability of the equations of motion for chain type interactions apply here. Namely, one requires that the cosmological constant contribution must be the same for each $W$-tensor:
\begin{align}
    \sum_{j=1}^{N-1} \Lambda_j^{(+)} &= \bar{\Lambda} \; ,
    \\
    \Lambda^{(-)}_j &= \bar{\Lambda} \; ,
\end{align}
defining $\Lambda_j^{(\pm)}$ as ($a_i^2/M_i^{D-2}$ times) the sums involving $\beta^{(0,j)}_m$ and $\beta_{D-m}^{(0,j)}$, respectively. The equations then reduce as before to $N$ copies of Einstein's equations for the $\bar{g}_{\mu\nu}$ common background, which one may in principle solve for $\bar{\Lambda}$ and the $N-1$ free conformal factors, after fixing one of them via coordinate rescaling.

For the non-proportional metric ansatze, the surviving off-diagonal parts of the $W$-tensors are also patterned as above, reflecting this new structure of the interactions. Explicitly, if one defines (C.F. Eqs. \eqref{sig_p} and \eqref{sig_m}):
\begin{align}
    \Sigma_{j}^{(+)} &= \sum_{m=0}^{D}\beta_m^{(0,j)}\binom{D-2}{m-1}a_{j}^{m-1}a_0^{1-m} \; ,
    \\
    \Sigma_{j}^{(-)} &= \sum_{m=0}^D\beta_{D-m}^{(0,j)}\binom{D-2}{m-1}a_0^{m-1}a_{j}^{1-m} \; ,
\end{align}
related as:
\begin{equation}
    \Sigma_{j}^{(-)} = \left(\frac{a_0}{a_{j}}\right)^{D-2}\Sigma_{j}^{(+)} \; ,
\end{equation}
then the star type analogue of (say) Eq. \eqref{W_nondiag} for the chain interactions is:
\begin{align}
    \begin{split}
        W^{(0)\mu}_{\;\;\;\;\;\;\nu} &= \frac{M_0^{D-2}\delta^\mu_\nu }{a_0^2}\sum_{j=1}^{N-1}\Lambda_j^{(+)}
        \\
        &+ \frac{l^\mu l_\nu}{2U} \sum_{j=1}^{N-1}\frac{a_{j}}{a_0}\left(r_{s,0}-r_{s,j}\right)\Sigma_j^{(+)}\; ,
    \end{split}
    \\
        W^{(j)\mu}_{\;\;\;\;\;\;\nu} &= \frac{M_j^{D-2}\delta^\mu_\nu }{a_j^2}\Lambda_j^{(-)} + \frac{l^\mu l_\nu}{2U} \frac{a_{0}}{a_j}\left(r_{s,j}-r_{s,0}\right)\Sigma_j^{(-)}\; ,
\end{align}
and similarly for the charged variants of the above (C.F. section \ref{Sec:RN_nonprop}). In principle, solutions then exist in each of the proportional (all $r_{s,j}=r_{s,0}$), non-proportional (all $\Sigma^{(+)}_j=0$) and partially proportional (combinations of both) branches.

As for the linearised equations, the star type analogue of Eq. \eqref{linearised eqs dg} for the metric perturbations is:
\begin{widetext}
\begin{align}
        \bar{\mathcal{E}}^{\alpha\beta}_{\;\mu\nu}\delta g^{(0)}_{\alpha\beta} + \bar{\Lambda}\delta g_{\mu\nu}^{(0)} + \frac{a_0^2}{M_0^{D-2}}\sum_{j=1}^{N-1}\Bigg\{\Sigma_j^{(+)}\left[\frac{a_{j}}{a_0}\left(\delta g^{(0)}_{\mu\nu}-\bar{g}_{\mu\nu}\delta g^{(0)}\right) - \frac{a_0}{a_{j}}\left(\delta g^{(j)}_{\mu\nu} - \bar{g}_{\mu\nu}\delta g^{(j)}\right)\right]\Bigg\} &= 0 \; ,
        \\
        \bar{\mathcal{E}}^{\alpha\beta}_{\;\mu\nu}\delta g^{(j)}_{\alpha\beta} + \bar{\Lambda}\delta g_{\mu\nu}^{(j)} + \frac{a_j^2}{M_j^{D-2}} \Sigma_j^{(-)}\left[\frac{a_{0}}{a_j}\left(\delta g^{(j)}_{\mu\nu}-\bar{g}_{\mu\nu}\delta g^{(j)}\right) - \frac{a_j}{a_{0}}\left(\delta g^{(0)}_{\mu\nu} - \bar{g}_{\mu\nu}\delta g^{(0)}\right)\right] &= 0 \; .
\end{align}
\end{widetext}
One may check, for example, that in $D=4$ spacetime dimensions, and denoting $a_0=1, a_j=c_j, \gamma_j=M_j/M_0$, these equations imply a mass matrix for the $\delta g^{(i)}_{\mu\nu}$ that takes the form:
\pagebreak
\begin{equation}
    \mathcal{M}^2 = \frac{1}{M_0^2}
    \begin{bmatrix}
        \sum_{j=1}^{N-1} \Sigma_j^{(+)} c_j & -\frac{\Sigma_1^{(+)}}{c_1} & -\frac{\Sigma_2^{(+)}}{c_2} & \hdots & -\frac{\Sigma_{N-1}^{(+)}}{c_{N-1}}
        \\
        -\frac{\Sigma_1^{(+)}c_1}{\gamma_1^2} & \frac{\Sigma_1^{(+)}}{c_1\gamma_1^2} & 0 & \hdots & 0
        \\
        -\frac{\Sigma_2^{(+)}c_2}{\gamma_2^2} & 0 & \frac{\Sigma_2^{(+)}}{c_2\gamma_2^2} & \hdots & 0
        \\
        \vdots & \vdots & \vdots & \ddots & \vdots
        \\
        -\frac{\Sigma_{N-1}^{(+)}c_{N-1}}{\gamma_{N-1}^2} & 0 & 0 & \hdots & \frac{\Sigma_{N-1}^{(+)}}{c_{N-1}\gamma_{N-1}^2}
    \end{bmatrix}
\end{equation}
This is precisely the star type interaction mass matrix that was derived for the $D=4$ multi-metric theory in \cite{prop_bg_multigrav}.

\bibliography{bibliography.bib}
\bibliographystyle{JHEP}
\end{document}